\documentclass[aps,prd,twocolumn,superscriptaddress,nofootinbib,floatfix]{revtex4-2}
\usepackage{amsmath,amssymb,bm}
\usepackage{graphicx}
\usepackage{float}
\usepackage{booktabs}
\usepackage{microtype}
\usepackage{xcolor}
\usepackage[colorlinks=true,linkcolor=blue!55!black,citecolor=blue!55!black,urlcolor=blue!55!black]{hyperref}
\usepackage[nameinlink,capitalise]{cleveref}
\crefname{figure}{Fig.}{Figs.}
\crefname{section}{Sec.}{Secs.}
\crefname{equation}{Eq.}{Eqs.}
\crefname{table}{Table}{Tables}
\Crefname{figure}{Fig.}{Figs.}
\Crefname{section}{Sec.}{Secs.}
\Crefname{equation}{Eq.}{Eqs.}
\Crefname{table}{Table}{Tables}
\newcommand{\Lsix}{\mathcal{L}_{6}^{v}}
\newcommand{\sigp}{\Sigma'}
\newcommand{\sigpp}{\Sigma''}
\begin{document}

\title{Nuclear interference versus dark sector excitation in the 248 keV LUX-ZEPLIN recoil candidate}

\author{Imtiaz Khan}
\email{ikhanphys1993@gmail.com}
\affiliation{Department of Physics, Zhejiang Normal University, Jinhua, Zhejiang 321004, China}
\affiliation{Research Center of Astrophysics and Cosmology, Khazar University, Baku, AZ1096, 41 Mehseti Street, Azerbaijan}

\author{Salvatore Capozziello}
\email{capozziello@na.infn.it}
\affiliation{Dipartimento di Fisica ``E. Pancini", Universit\`a di Napoli ``Federico II", Complesso Universitario di Monte Sant’ Angelo, Edificio G, Via Cinthia, I-80126, Napoli, Italy,}
\affiliation{Istituto Nazionale di Fisica Nucleare (INFN), sez. di Napoli, Via Cinthia 9, I-80126 Napoli, Italy,}
\affiliation{Scuola Superiore Meridionale, Via Mezzocannone 4, I-80134, Napoli, Italy.}

\author{G. Mustafa}
\email{gmustafa3828@gmail.com}
\affiliation{Department of Physics, Zhejiang Normal University, Jinhua, Zhejiang 321004, China}

\author{Farruh~Atamurotov}
\email{atamurotov@yahoo.com}
\affiliation{School of Physics, Harbin Institute of Technology, Harbin 150001, People's Republic of China}
\affiliation{University of Tashkent for Applied Sciences, Str. Gavhar 1, Tashkent 100149, Uzbekistan}
\affiliation{National University of Uzbekistan, Tashkent 100174, Uzbekistan}

\author{Ahmadjon~Abdujabbarov}
\email{ahmadjonab@gmail.com}
\affiliation{School of Physics, Harbin Institute of Technology, Harbin 150001, People's Republic of China}

\author{Chengxun Yuan}
   \email{yuancx@hit.edu.cn}
\affiliation{School of Physics, Harbin Institute of Technology, Harbin 150001, People's Republic of China}

\begin{abstract}
The 2026 LUX-ZEPLIN (LZ) high energy nuclear recoil search reports one event at $248\pm23_{\rm stat}\pm23_{\rm sys}~\mathrm{keV}$ and finds the elastic isovector interaction $\mathcal L_6^v$ locally favored at about $3.3\sigma$ in the heavy dark matter regime. We examine whether this recoil scale can arise from resolved nuclear response within the elastic interaction and compare its spectral, target, and annual modulation behavior with endothermic dark sector excitation. Covariant matching correlates four Galilean operators, removes the longitudinal spin response algebraically, and fixes interference between density and orbital spin orbit amplitudes. The GCN and JJ55 xenon shell model calculations give a second natural xenon cancellation at $211$ and $214~\mathrm{keV}$. In the heavy mass regime, its position changes by less than $0.01~\mathrm{keV}$ from a dark matter mass of $200~\mathrm{GeV}$ to the asymptotic limit. A curvature weighted isotope centroid reproduces the position and residual depth of the natural xenon minimum. Direct $^{40}$Ar calculations shift the nuclear feature to about $350$ and $425~\mathrm{keV}$, whereas endothermic excitation follows reduced mass scaling set by the dark state splitting. At a dark matter mass of $1~\mathrm{TeV}$, the endothermic Ar to Xe scale ratio is $1.082$, compared with $1.63$ to $2.01$ for the nuclear calculations. Elastic scattering gives a few percent annual modulation in the adopted halo model, while endothermic solutions near the maximum laboratory halo speed show much larger seasonal variation. These scaling and timing behaviors provide tests of a target dependent nuclear cancellation against an excitation energy set by dark sector kinematics.
\end{abstract}

\maketitle

\section{Introduction}
\label{secintro}

High energy direct detection probes momentum transfers at which finite size nuclear structure and dark sector kinematics can both affect the recoil spectrum. The latest 2026 LUX-ZEPLIN result uses $2.84~\mathrm{tonne\,yr}$ in a nuclear recoil window extending to about $270~\mathrm{keV}$ and reports one event consistent with a $248\pm23_{\rm stat}\pm23_{\rm sys}~\mathrm{keV}$ nuclear recoil in a low background region. The background only hypothesis has a global tension of $2.6\sigma$ after the look elsewhere correction, and several interactions have local significances above $3\sigma$~\cite{LZ2026highE}. The question is which part of the scattering dynamics sets this recoil scale. A dark sector excitation and a finite momentum nuclear cancellation can occur in the same energy range, but their mass, target, spectral, and annual modulation dependences differ.

One possibility is that the recoil scale is associated with dark sector spectroscopy. In endothermic scattering, an internal splitting modifies threshold kinematics, target response, and annual modulation~\cite{TuckerSmith2001hy,TuckerSmith2004jv,Chang2009idm}, while model independent formulations make the target and velocity dependence explicit~\cite{Barello2014eaa}. High recoil studies consider splittings of a few hundred keV~\cite{DiMauro2026lz,Su2026lz,Bramante2016rdh}, including dark photon~\cite{Yamashita2026lz}, Higgsino~\cite{FanReece2026lz,Freese2026lz,WuZhangZhu2026lz}, and broader electroweak multiplet realizations~\cite{Smirnov2026lz}. Peccei Quinn constructions relate comparable splittings to mixed axion and electroweak dark matter~\cite{Yin2026lz,Visinelli2026lz}. In these cases, the characteristic recoil energy is tied to an internal dark sector scale.

A Higgsino splitting of a few hundred keV also constrains physics well above the recoil scale. Its ultraviolet interpretation depends on the neutralino mixing pattern and supersymmetric boundary conditions. Generic electroweak mixing can place gauginos many orders of magnitude above the Higgsino mass~\cite{FanReece2026lz,Yin2026lz}, whereas nonuniversal gaugino relations allow much lower electroweak gaugino scales for a comparable splitting~\cite{DuWang2026lz}. Solar capture supplies an independent constraint for a full density thermal Higgsino under the assumptions of Ref.~\cite{PospelovRamani2026lz}. The high energy sideband gives another test, with sensitivity controlled by experimental acceptance~\cite{Rodd2026sideband}. Inferring gaugino masses from a recoil therefore requires first establishing a neutral state splitting and then specifying the electroweak mixing structure.

Other proposals associate the recoil energy with different parts of the scattering process. Fermionic absorption converts dark matter rest energy into nuclear recoil and faces an independent scintillator constraint~\cite{Lou2026lz}. Atmospheric neutrino upscattering produces a massive final state through a different reaction~\cite{Jeesun2026nu}. An axion portal realization of elastic pseudoscalar scattering uses the momentum dependence of another LZ tested interaction~\cite{Unwin2026lz}. Endothermic analyses also find large annual modulation when the required speed approaches the maximum laboratory halo speed~\cite{McCabe2026seasonal}. Comparing these mechanisms therefore requires the recoil energy together with the neighboring spectral shape, target dependence, and annual modulation.

At $248~\mathrm{keV}$, the momentum transfer in xenon is about $0.25~\mathrm{GeV}$, corresponding to a wavelength comparable to nuclear length scales. The nonrelativistic effective field theory (NREFT) of dark matter scattering contains density, spin, orbital, and spin orbit responses absent from a point nucleus description~\cite{Fan2010gt,Fitzpatrick2012ix,Anand2013yka}. Operator interference modifies the relation between recoil spectra and particle couplings~\cite{Cirelli2013tools,Gresham2014vja,Catena2014uqa}. Chiral effective field theory (chiral EFT) organizes one body and two body currents~\cite{Hoferichter2015chiral,Hoferichter2016analysis}, while shell model calculations resolve xenon spin and orbital responses~\cite{Menendez2012tm,Klos2013rwa,Vietze2014jga}. Recent many body studies extend finite momentum xenon calculations~\cite{Hoferichter2018acd,AbdelKhaleq2022natXe,AbdelKhaleq2024nuc}. Liquid xenon EFT searches use enlarged recoil windows for interactions not concentrated near threshold~\cite{LUX2021eft,XENON2024eft,LZ2024nreft}, supported by recoil window optimization studies~\cite{Bozorgnia2018dvr}. The covariant LZ basis further imposes definite relations among nonrelativistic coefficients~\cite{LZ2024covariant}. We follow these relations through the nuclear amplitudes and evaluate the cancellation energy, isotope dependence, target dependence, and annual modulation for one elastic interaction.

We consider the elastic isovector interaction $\Lsix$ already tested in the LZ covariant basis. Its local significance increases from $2.9\sigma$ at $100~\mathrm{GeV}$ to about $3.3\sigma$ in the TeV region~\cite{LZ2026highE}. The heavy mass plateau allows the kinematic mass dependence to be separated from the finite momentum nuclear response. \Cref{seclzmeaning} discusses this loss of mass sensitivity, \cref{secresponse} derives the correlated nuclear response, and \cref{secxenon,sectargets,sectime} examine xenon, target, and annual modulation tests. The covariant coefficient relation selects a cancellation between density and orbital spin orbit amplitudes. Its mass dependence, isotope composition, target shift, and timing behavior can then be compared with those of an endothermic excitation scale.

\begin{figure}[t]
\centering
\includegraphics[width=\columnwidth]{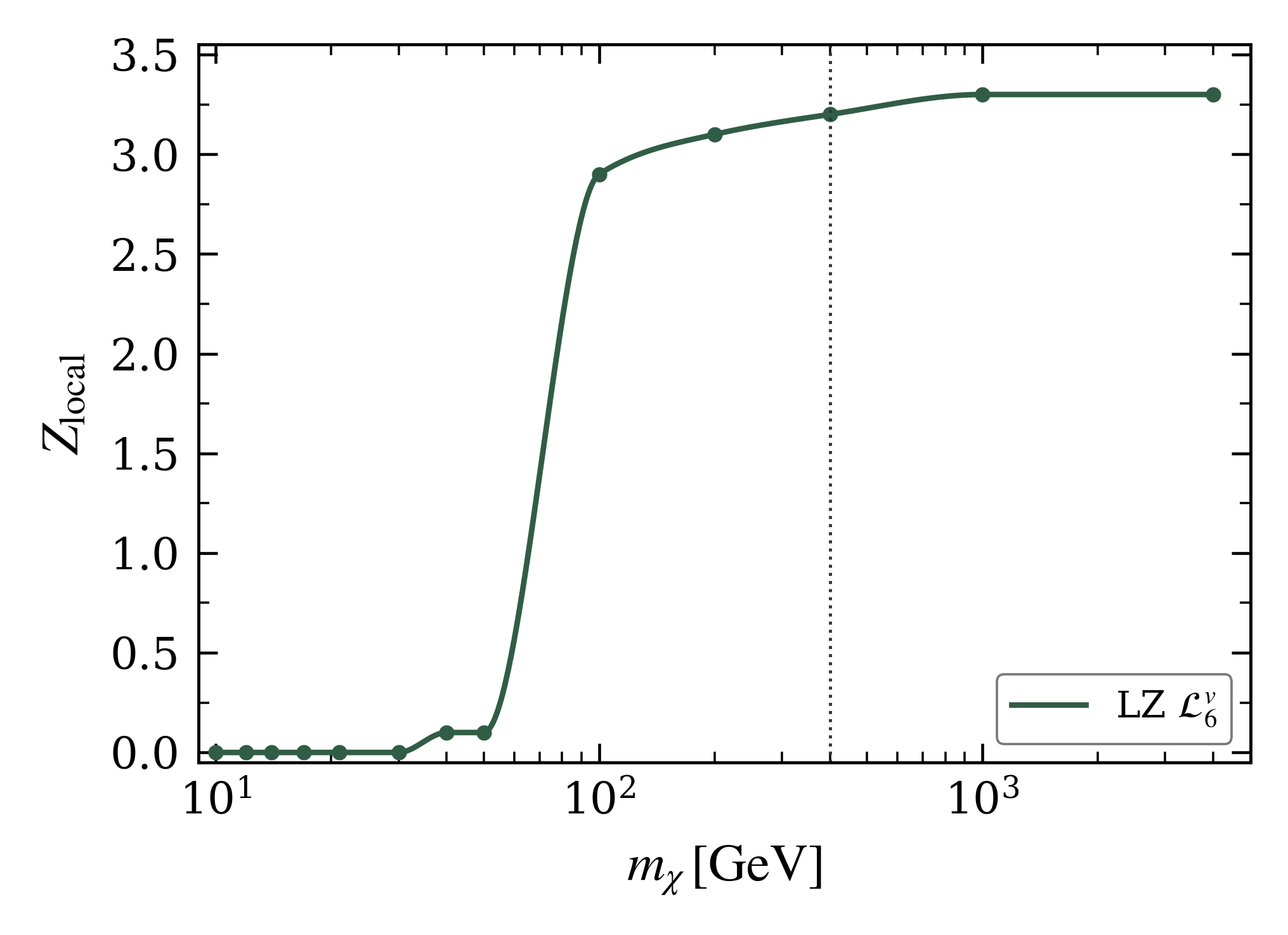}
\caption{Local significance reported by LZ for the isovector $\mathcal L_6^v$ interaction from Table S7 of Ref.~\cite{LZ2026highE}. Markers give the tabulated values and the continuous curve is a piecewise cubic Hermite interpolation through them. The heavy mass plateau develops as the dark matter xenon reduced mass approaches the nuclear mass.}
\label{figlzsig}
\end{figure}

\section{Mass saturation and recoil scale}
\label{seclzmeaning}

The first comparison concerns mass dependence. \Cref{figlzsig} shows the heavy mass plateau reported for $\Lsix$. Let $E_R$ denote nuclear recoil energy, $m_\chi$ the dark matter mass, $m_A$ the target nucleus mass, and $v_{\min}$ the minimum incident speed required for the recoil. We use natural units in the kinematic expressions, quote masses in GeV and recoil energies in keV, and give numerical speeds in $\mathrm{km\,s^{-1}}$. For elastic scattering~\cite{Lewin1995rx,Baxter2021pqo}
\begin{equation}
 v_{\min}(E_R)=\sqrt{\frac{m_AE_R}{2\mu_{\chi A}^2}},
 \qquad
 \mu_{\chi A}=\frac{m_\chi m_A}{m_\chi+m_A},
\label{eqvmin}
\end{equation}
where $\mu_{\chi A}$ is the dark matter nucleus reduced mass. In the limit $m_\chi\gg m_A$, \cref{eqvmin} approaches $v_{\min}=\sqrt{E_R/(2m_A)}$, so the required speed becomes only weakly dependent on $m_\chi$. The remaining mass sensitivity is measured by
\begin{equation}
\frac{\partial\ln v_{\min}}{\partial\ln m_\chi}
=-\frac{m_A}{m_\chi+m_A}.
\label{eqmasssaturation}
\end{equation}
For representative $^{131}$Xe, the magnitude decreases from $0.234$ at $m_\chi=400~\mathrm{GeV}$ to $0.109$ at $1~\mathrm{TeV}$ and $0.0296$ at $4~\mathrm{TeV}$. Equation~\cref{eqmasssaturation} quantifies the loss of kinematic mass sensitivity across the LZ $\mathcal L_6^v$ plateau. A recoil feature with a stable position can therefore remain sensitive to the interaction and nuclear matrix elements after the dark matter mass dependence has nearly saturated.

The reconstructed nuclear cancellation behaves differently. Scanning the full natural xenon rate from $m_\chi=200~\mathrm{GeV}$ through the heavy regime places the high GCN minimum between $211.066$ and $211.073~\mathrm{keV}$ and the JJ55 minimum between $214.022$ and $214.024~\mathrm{keV}$. The displacement remains below $0.01~\mathrm{keV}$. Kinematic weighting changes the depth more strongly than the position. For $\Lsix$, the high energy minimum is therefore set mainly by the nuclear amplitude and has negligible heavy mass drift.

Endothermic scattering has a different mass dependence. A positive dark state splitting $\delta$, defined as the final state mass minus the initial state mass, adds an excitation term to the scattering kinematics~\cite{TuckerSmith2004jv,Barello2014eaa}. The corresponding minimum speed is
\begin{equation}
 v_{\min}^{\rm endo}(E_R)=\frac{1}{\sqrt{2m_AE_R}}
 \left(\frac{m_AE_R}{\mu_{\chi A}}+\delta\right).
\label{eqvinel}
\end{equation}
The additional term shifts the preferred recoil scale toward a value controlled by $\delta$ and increases the incident speed required to populate the recoil. As a result, an elastic nuclear cancellation and an endothermic excitation may occur in the same recoil interval while following different target and timing dependences.

\section{Covariant interaction and nuclear cancellation}
\label{secresponse}

A mass stable recoil feature requires a finite momentum structure in the scattering amplitude. The Galilean response formalism organizes direct detection amplitudes in a general operator basis~\cite{Fitzpatrick2012ib,Fitzpatrick2012ix,Anand2013yka}. In the covariant basis used by LZ, the interaction is~\cite{LZ2024covariant}
\begin{equation}
\mathcal L_6=d_6\,\bar\chi\gamma^\mu\chi\,
\bar N i\sigma_{\mu\alpha}\frac{q^\alpha}{m_N}N,
\label{eqL6}
\end{equation}
Here $\chi$ denotes a spin $1/2$ dark matter field, $N$ the nucleon field, $q^\mu$ the four momentum transfer, $m_N$ the nucleon mass, and $d_6$ the dimensionless coupling in the LZ convention. The matrices satisfy $\sigma_{\mu\alpha}=i[\gamma_\mu,\gamma_\alpha]/2$. The superscript $v$ specifies the isovector nucleon combination, with opposite proton and neutron signs in the convention adopted here. The nonrelativistic reduction maps this interaction onto the Galilean operators according to
\begin{multline}
\mathcal L_6\rightarrow d_6\left[
\frac{x}{2}\mathcal O_1-2\mathcal O_3\right.\\
\left.+2\frac{m_N}{m_\chi}\left(x\mathcal O_4-\mathcal O_6\right)
\right],
\label{eqnrmap}
\end{multline}
In \cref{eqnrmap}, $q=|\mathbf q|$ and $x\equiv q^2/m_N^2$. In the standard Galilean basis, $\mathcal O_1$ is the spin independent density operator, $\mathcal O_3=i\mathbf S_N\cdot(\mathbf q/m_N\times\mathbf v^\perp)$ is the nucleon spin orbit operator, $\mathcal O_4=\mathbf S_\chi\cdot\mathbf S_N$ is the spin spin operator, and $\mathcal O_6=(\mathbf S_\chi\cdot\mathbf q/m_N)(\mathbf S_N\cdot\mathbf q/m_N)$ is the longitudinal spin momentum operator~\cite{Fan2010gt,Fitzpatrick2012ix,Anand2013yka}. The vectors $\mathbf S_\chi$ and $\mathbf S_N$ are the dark matter and nucleon spins, and $\mathbf v^\perp$ is the relative velocity transverse to $\mathbf q$. The covariant reduction therefore fixes the four coefficients as
\begin{align}
c_1&=\frac{x}{2}d_6, & c_3&=-2d_6,\nonumber\\
c_4&=2d_6\frac{m_N}{m_\chi}x, & c_6&=-2d_6\frac{m_N}{m_\chi}.
\label{eqcoeff}
\end{align}
In a general NREFT treatment these coefficients may be varied independently. For the covariant interaction, however, the relations in \cref{eqcoeff} fix their relative magnitudes and signs before the nuclear wave function is specified. This correlation is the origin of the response structure considered below.

The functions $R_X^{11}$ contain the particle physics coefficients for nuclear channel $X$, with the superscript $11$ denoting the isovector isovector component. The responses $\Sigma''$ and $\Sigma'$ describe nucleon spin components longitudinal and transverse to the momentum transfer. For spin $1/2$ dark matter, $\Sigma''$ depends on the combination $c_4+xc_6$~\cite{Anand2013yka}. Substitution of \cref{eqcoeff} gives $c_4+xc_6=0$, so $R_{\sigpp}^{11}=0$. This algebraic cancellation occurs before the nuclear matrix elements are evaluated and removes the longitudinal spin contribution for every target.

The transverse spin response remains,
\begin{equation}
R_{\sigp}^{11}=d_6^2\left[
\frac{x}{2}v_T^{\perp2}
+\frac14\frac{m_N^2}{m_\chi^2}x^2
\right],
\label{eqsigp}
\end{equation}
where $v_T^\perp$ is the transverse relative speed at nuclear level. In \cref{eqsigp}, the first term is velocity suppressed and the second decreases as $m_\chi^{-2}$ in the heavy regime. The high recoil structure is therefore controlled mainly by the scalar density and orbital spin orbit channels.

The leading scalar sector contains the density response $M$, orbital spin orbit response $\Phi''$, and interference response $\Phi''M$. Using the convention of Ref.~\cite{Anand2013yka}, the $\mathcal O_1$ and $\mathcal O_3$ coefficients give
\begin{align}
R_M^{11}&=\frac14d_6^2x^2, &
 xR_{\Phi''}^{11}&=d_6^2x^2,\nonumber\\
 xR_{\Phi''M}^{11}&=-d_6^2x^2.
\label{eqscalarparts}
\end{align}
All three terms contain the same $q^4$ factor, while their relative signs are fixed by the covariant matching. Combining these coefficients with the nuclear response functions $W_X^{11}$, we write the scalar target response as $P_{6,{\rm scalar}}^v$,
\begin{equation}
P_{6,{\rm scalar}}^v=d_6^2x^2
\left[
\frac14W_M^{11}+W_{\Phi''}^{11}-W_{\Phi''M}^{11}
\right].
\label{eqscalarresponse}
\end{equation}
The $q^4$ prefactor enhances high recoil energies, while the negative interference term can suppress the scalar response at selected momentum transfer. The cancellation position is therefore determined by the relative density and orbital spin orbit amplitudes rather than by the momentum prefactor alone.

\begin{figure}[t]
\centering
\includegraphics[width=\columnwidth]{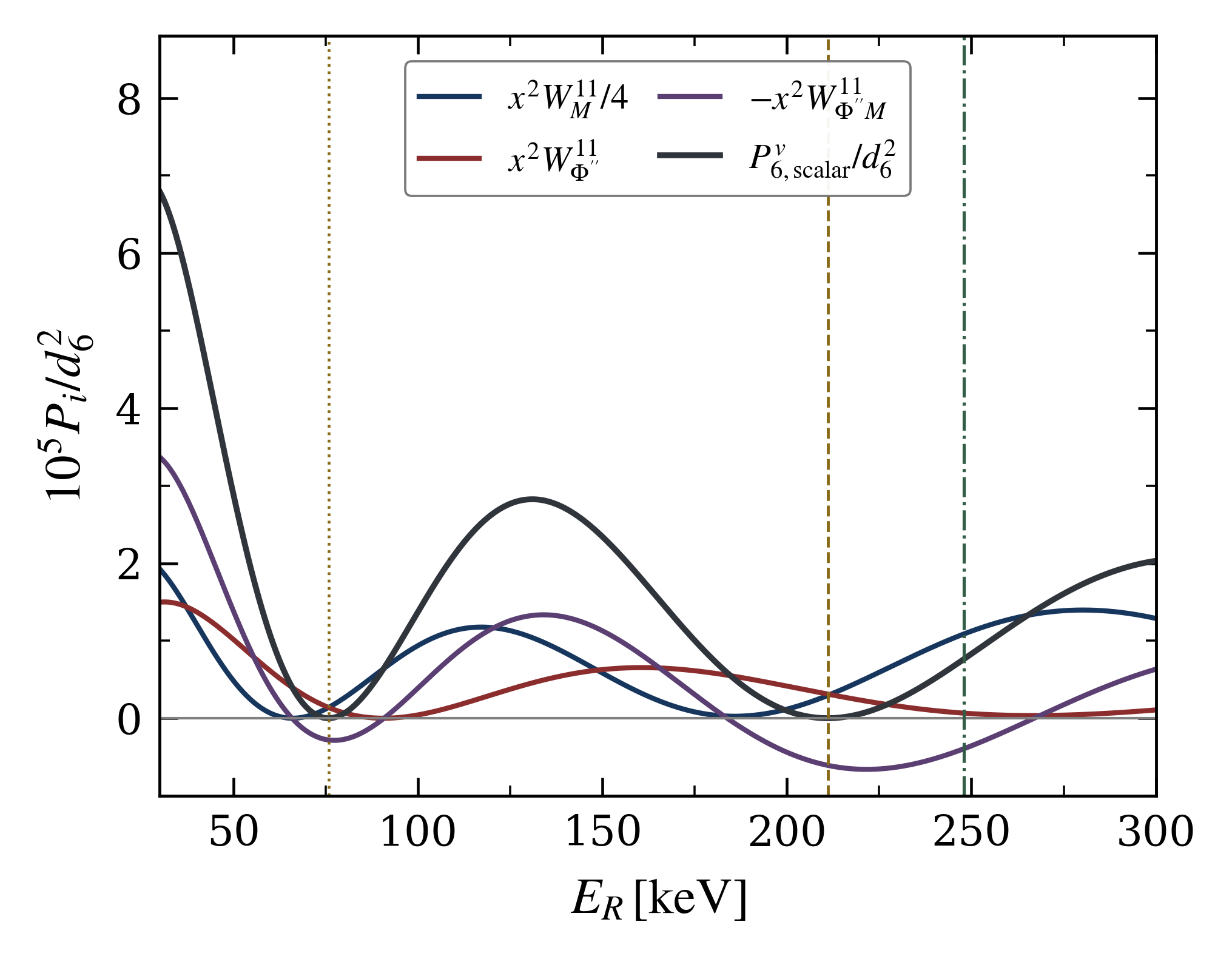}
\caption{Natural xenon scalar response from the GCN one body density matrices. Density, spin orbit, and interference contributions follow \cref{eqscalarresponse}. Their correlated sum develops cancellations near $76$ and $211~\mathrm{keV}$. The green line marks the reported LZ recoil energy~\cite{LZ2026highE}.}
\label{figresponses}
\end{figure}

The amplitude structure is clearer for isotope fraction $f_A$, nuclear spin $J_A$, and multipole rank $J$. Let $\mathcal M_{M,J}^{(1)}$ and $\mathcal M_{\Phi'',J}^{(1)}$ denote the isovector reduced one body amplitudes in the response convention~\cite{Gorton2022cpc}. Then
\begin{equation}
\frac{P_{6,{\rm scalar}}^v}{d_6^2x^2}
=\sum_{A,J}\frac{f_A}{2J_A+1}
\left(
\frac12\mathcal M_{M,J}^{(1)}-\mathcal M_{\Phi'',J}^{(1)}
\right)^2.
\label{eqamplitudesquare}
\end{equation}
This is a positive sum over isotope and multipole contributions. Since isotope amplitudes are squared before being added, a deep natural xenon minimum requires several amplitudes to approach zero over a common momentum interval. The GCN decomposition in \cref{figresponses} shows the density, orbital spin orbit, and interference terms whose correlated sum produces the two cancellations.

\section{Resolved xenon structure}
\label{secxenon}

GCN and JJ55 are two shell model Hamiltonians used to construct xenon one body density matrices. Their valence space contains the orbitals $0g_{7/2}$, $1d_{5/2}$, $1d_{3/2}$, $2s_{1/2}$, and $0h_{11/2}$~\cite{Gorton2022cpc}. The $\Phi''$ operator probes orbital spin orbit structure and can remain finite for a nucleus with ground state spin $J_A=0$. Consequently, the cancellation is controlled by orbital occupancies and multipole matrix elements rather than by the ground state spin quantum number alone.

LZ evaluates its signal spectra with WimPyDD and one body density matrices developed for DMFormFactor v6, including the nuclear updates described in the 2024 NREFT study~\cite{LZ2026highE,LZ2024nreft}. We evaluate the finite momentum response with the public GCN and JJ55 matrices distributed with the response calculation~\cite{Gorton2022cpc}. The quoted node positions and target shifts therefore refer to these matrix sets. The coefficient relations in \cref{eqcoeff} and the positive amplitude square in \cref{eqamplitudesquare} follow from the interaction and hold independently of the xenon Hamiltonian, whereas the numerical cancellation energies depend on the one body density matrices.

The shell model matrix elements are represented in a harmonic oscillator basis. For nuclear mass number $A$, the Blomqvist Molinari oscillator length $b$ and dimensionless momentum variable $y$ are~\cite{Gorton2022cpc}
\begin{equation}
b^2=\frac{41.467}{45A^{-1/3}-25A^{-2/3}}~\mathrm{fm}^2,
\qquad
y=\left(\frac{qb}{2\hbar c}\right)^2.
\label{eqoscillator}
\end{equation}
We use $\hbar c=0.197327~\mathrm{GeV\,fm}$ in \cref{eqoscillator} to combine momentum in GeV with $b$ in fm. For representative $^{131}$Xe, $b=2.29~\mathrm{fm}$. The high energy GCN and JJ55 minima occur at $y\simeq1.75$ and $1.78$, while the $248~\mathrm{keV}$ recoil corresponds to $y\simeq2.06$. Values of order unity show that finite nuclear size and orbital structure are relevant beyond the coherent long wavelength limit.

We denote the first and second cancellation energies by $E_{\star,1}$ and $E_{\star,2}$. The continuous response gives
\begin{align}
E_{\star,1}^{\rm GCN}&=75.8~\mathrm{keV}, & E_{\star,2}^{\rm GCN}&=211~\mathrm{keV},\nonumber\\
E_{\star,1}^{\rm JJ55}&=78.4~\mathrm{keV}, & E_{\star,2}^{\rm JJ55}&=214~\mathrm{keV}.
\label{eqtwonodes}
\end{align}
The reported recoil therefore lies above the second minimum for both Hamiltonians, on the part of the spectrum that rises as the amplitude moves away from the zero. The lower minimum is generated by the same density and orbital spin orbit interference. Thus, the elastic interaction contains two occurrences of the same cancellation mechanism, with the lower one providing an additional spectral feature at smaller recoil energy.

Shell model calculations separate the coherent density channel from spin and orbital responses at finite momentum~\cite{Menendez2012tm,Klos2013rwa,Vietze2014jga}. Chiral EFT places these responses in a matching framework containing coherent and two body currents~\cite{Hoferichter2015chiral,Hoferichter2016analysis,Hoferichter2018acd}. Recent xenon and broader nuclear calculations further show the target dependence of nonstandard responses~\cite{AbdelKhaleq2022natXe,AbdelKhaleq2024nuc,AbdelKhaleq2025cevns}. Here GCN and JJ55 provide a two Hamiltonian comparison of the second cancellation. Their separation of about $3~\mathrm{keV}$ shows that the high node persists under this change of nuclear interaction, but is not a complete nuclear uncertainty. Such an estimate requires additional Hamiltonians with consistently recomputed one body densities.

Let $R(E_R)=dR/dE_R$ denote the differential recoil rate. At $m_\chi=1~\mathrm{TeV}$, the normalized ratios are $R(248)/R(50)=0.112$ for GCN and $0.0721$ for JJ55. At the second minima, the transverse spin fractions are $1.83\times10^{-4}$ and $5.07\times10^{-4}$, and both fall below $5.2\times10^{-7}$ at $248~\mathrm{keV}$. The scalar density and orbital spin orbit sector therefore controls the dip and subsequent high energy rise.

\begin{figure}[t]
\centering
\includegraphics[width=\columnwidth]{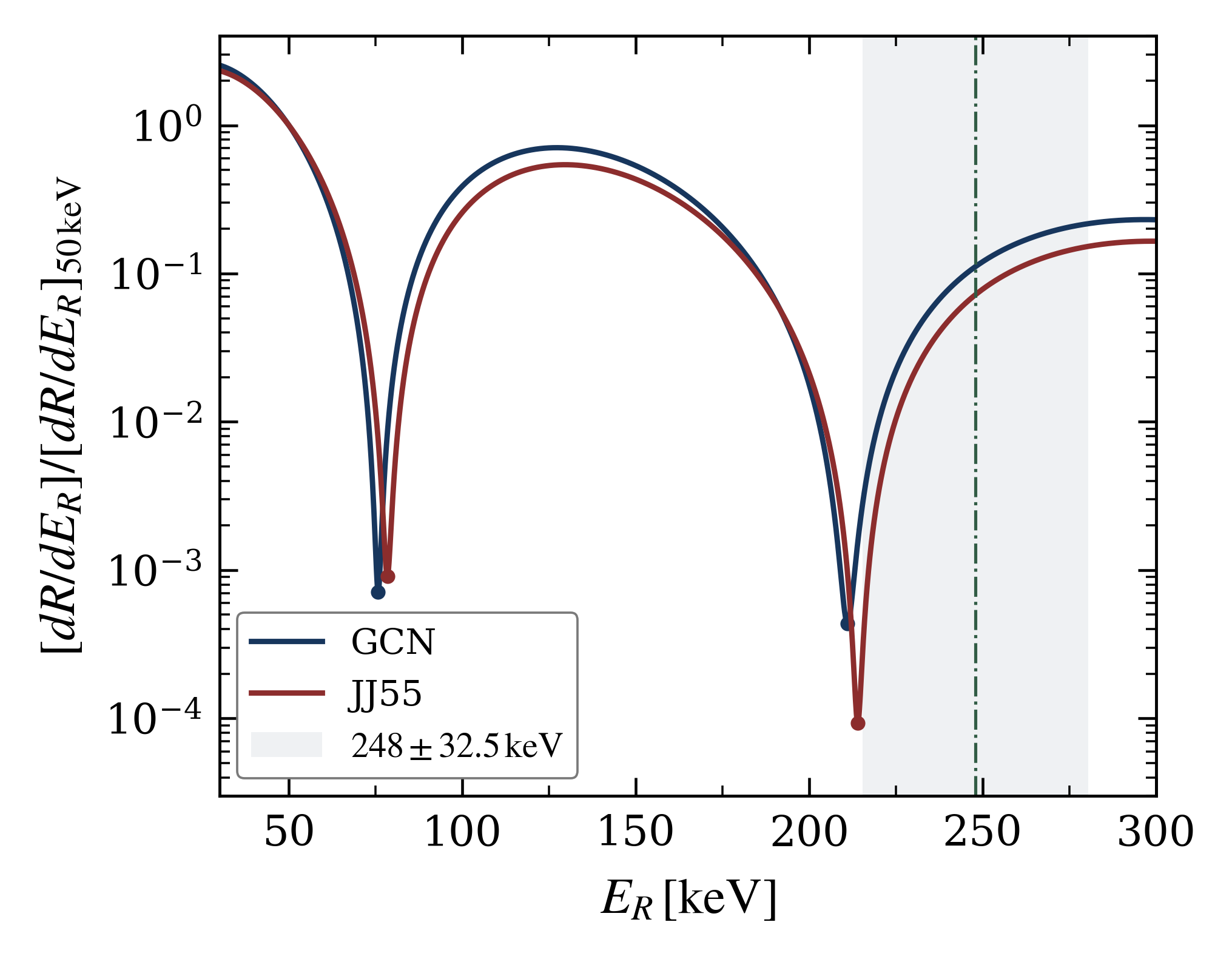}
\caption{Normalized $\mathcal L_6^v$ recoil spectrum at $m_\chi=1~\mathrm{TeV}$ for GCN and JJ55, including the surviving $\Sigma'$ response. Both finite momentum minima are visible. The shaded interval is $248\pm32.5~\mathrm{keV}$ from quadrature combination of the quoted statistical and systematic recoil energy uncertainties~\cite{LZ2026highE}.}
\label{figspectrum}
\end{figure}

The spectrum above the second node provides a further test. At $1~\mathrm{TeV}$, $R(270)/R(248)=1.73$ for GCN and $1.85$ for JJ55. Broad maxima occur near $297$ and $299~\mathrm{keV}$, where the rates reach $2.07$ and $2.29$ times their values at $248~\mathrm{keV}$. The spectra in \cref{figspectrum} show both cancellations and the subsequent rise. LZ reports high nuclear recoil efficiency through $250~\mathrm{keV}$ and an efficiency roll off near $270~\mathrm{keV}$~\cite{LZ2026highE}. The present window therefore covers part of the rising spectrum. Additional exposure can constrain the local slope, while a calibrated extension above the roll off would probe the approach to the broad maximum.

Combining the statistical and systematic recoil energy uncertainties in quadrature gives $\sigma_E=32.5~\mathrm{keV}$. The GCN and JJ55 minima lie $1.14\sigma_E$ and $1.05\sigma_E$ below the central recoil energy. The lower edge of the combined interval is $215.5~\mathrm{keV}$, only $4.4~\mathrm{keV}$ above GCN and $1.5~\mathrm{keV}$ above JJ55. The present energy uncertainty therefore does not resolve the minimum from the neighboring rising spectrum. A larger high energy recoil sample can test this spectral shape.

\subsection{Isotope composition}

\begin{figure}[t]
\centering
\includegraphics[width=\columnwidth]{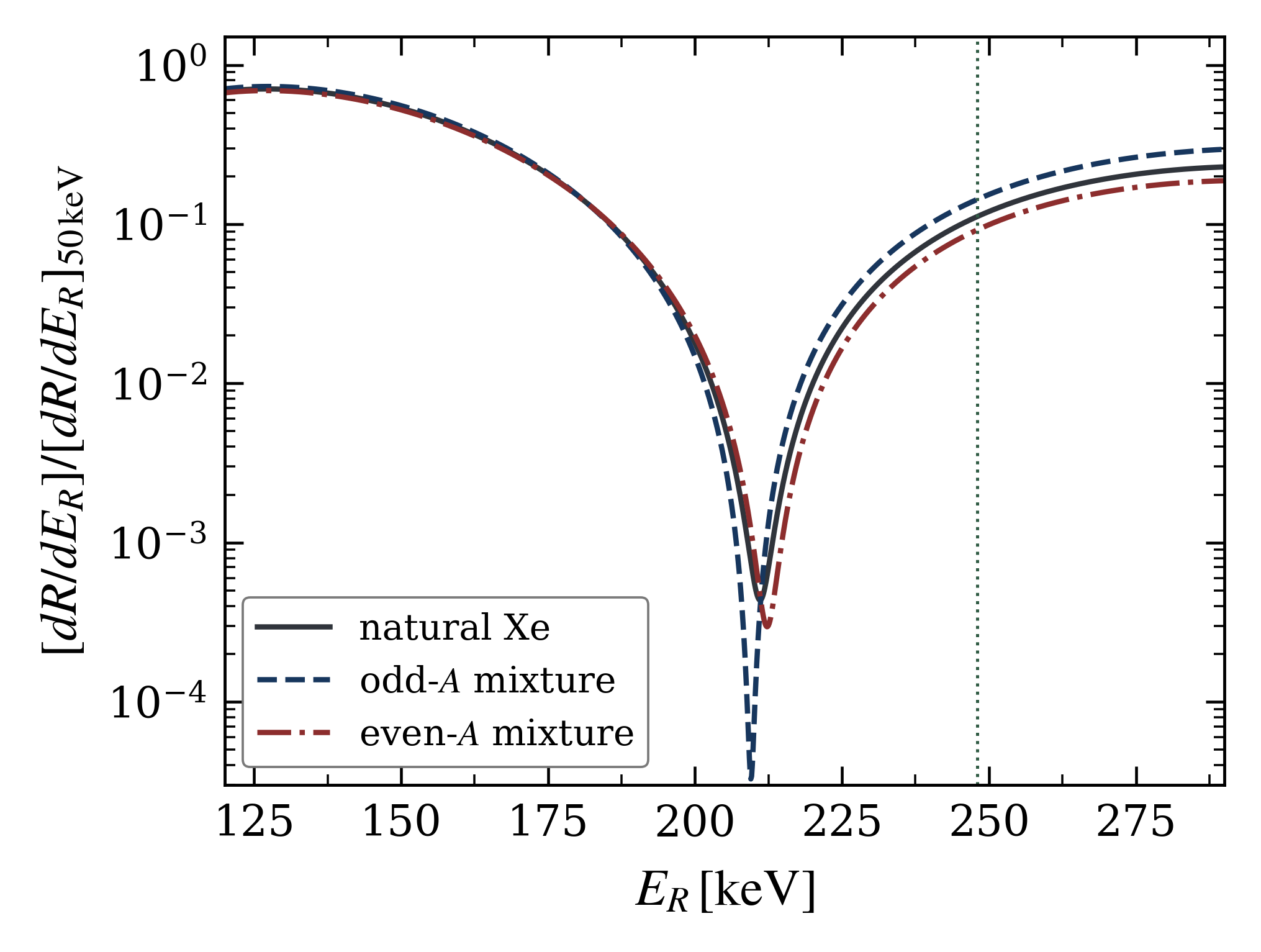}
\caption{GCN recoil shapes for natural xenon and abundance normalized odd $A$ and even $A$ isotope mixtures. Both isotope classes retain a high energy cancellation. The persistence in even isotopes supports the orbital origin of the $\Phi''$ structure.}
\label{figisotopes}
\end{figure}

The isotope decomposition shows the many body origin of the natural xenon minimum. Even and odd xenon isotopes contribute to the orbital spin orbit response. At $m_\chi=1~\mathrm{TeV}$, the individual GCN minima range from $208.03$ to $214.23~\mathrm{keV}$, while JJ55 ranges from $212.89$ to $217.32~\mathrm{keV}$. Since \cref{eqamplitudesquare} is a positive isotope and multipole sum, the natural xenon minimum results from the near alignment of isotope specific amplitude zeros. The GCN mixtures in \cref{figisotopes} show the cancellation in both even and odd isotope classes.

The nearby isotope minima can be described analytically. Let $R_A(E_R)$ be the differential rate of isotope $A$, let $E_A$ denote its local minimum, and define $R_{A,0}=R_A(E_A)$. In the neighborhood of $E_A$ we expand $R_A(E_R)=R_{A,0}+\kappa_A(E_R-E_A)^2+\cdots$, where $\kappa_A=R_A''(E_A)/2$. With the natural abundance and spin weight $\omega_A=f_A/(2J_A+1)$ and $K=\sum_A\omega_A\kappa_A$, quadratic minimization gives
\begin{align}
E_\star^{(2)}&=K^{-1}\sum_A\omega_A\kappa_AE_A,\label{eqisotopecentroid}\\
R_{\rm Xe}(E_\star^{(2)})&=\sum_A\omega_AR_{A,0}+\sum_A\omega_A\kappa_A(E_A-E_\star^{(2)})^2.\label{eqisotopedepth}
\end{align}
Equation~\cref{eqisotopecentroid} weights the isotope minima by both their abundance and their local curvature, whereas \cref{eqisotopedepth} relates the residual natural xenon rate to the weighted spread of those minima. At $1~\mathrm{TeV}$, the centroid gives $211.018~\mathrm{keV}$ for GCN, compared with the full minimum at $211.068~\mathrm{keV}$, and $214.009~\mathrm{keV}$ for JJ55, compared with $214.023~\mathrm{keV}$. The corresponding quadratic depths differ from the full rates by $1.35\%$ and $0.602\%$. The local expansion therefore reproduces the natural target minima to better than $0.1~\mathrm{keV}$ and shows how imperfect alignment of the isotope zeros leaves a finite residual rate.

Isotopic enrichment changes the isotope weights and can therefore shift and reshape the cancellation. The shift depends on orbital occupancies and multipole matrix elements beyond the ground state spin label. With sufficient high recoil exposure, enriched xenon provides an additional nuclear structure test.

\section{Target dependence}
\label{sectargets}

\begin{figure*}[t]
\centering
\includegraphics[width=0.48\textwidth]{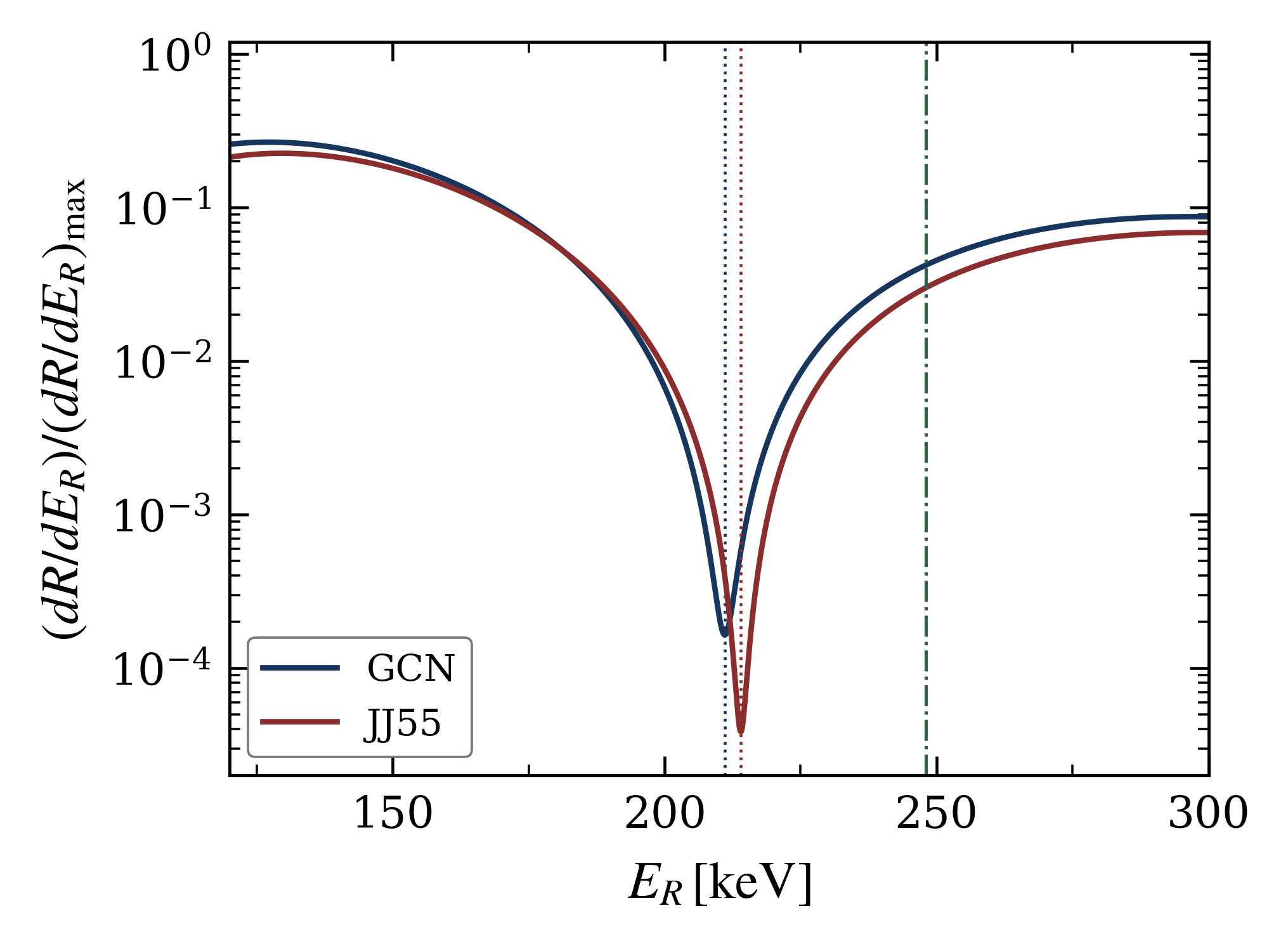}\hfill
\includegraphics[width=0.48\textwidth]{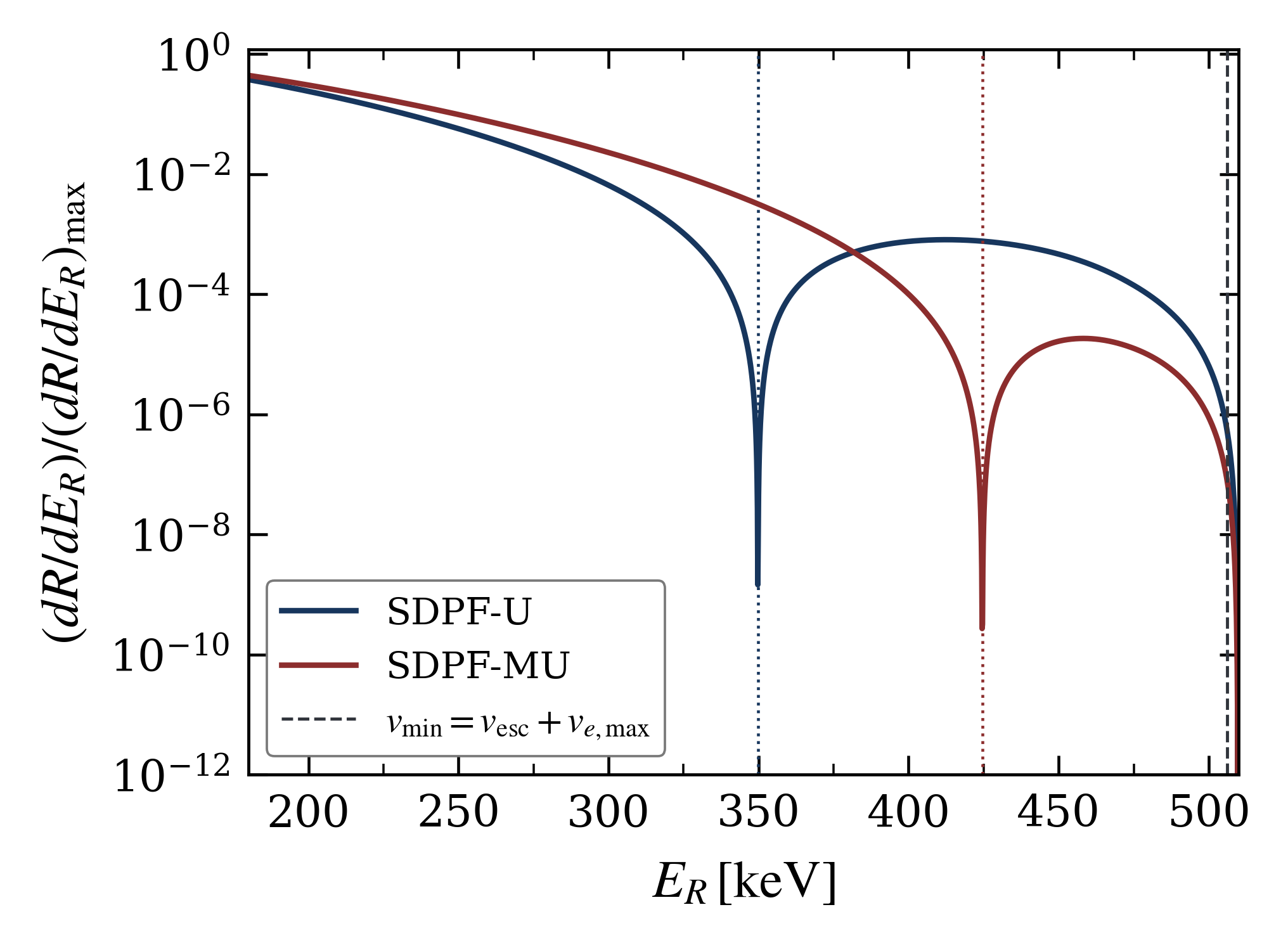}
\caption{Target dependence of the high energy $\mathcal L_6^v$ response. The left panel shows natural xenon with the second GCN and JJ55 cancellations near $211$ to $214~\mathrm{keV}$. The right panel shows $^{40}$Ar with SDPF U and SDPF MU cancellations at substantially larger recoil energy. The dashed argon line marks the approximate $1~\mathrm{TeV}$ elastic kinematic cutoff under the adopted halo prescription.}
\label{figtargets}
\end{figure*}

A second target tests whether the recoil scale follows nuclear structure or reduced mass kinematics. We repeat the scalar $\mathcal L_6^v$ contraction for $^{40}$Ar using the SDPF U and SDPF MU shell model interactions distributed with the response calculation~\cite{Gorton2022cpc}. Both Hamiltonians give a finite $\Phi''$ response and a high energy cancellation for the $0^+$ argon ground state. At $m_\chi=1~\mathrm{TeV}$, the minima occur near $350~\mathrm{keV}$ for SDPF U and $425~\mathrm{keV}$ for SDPF MU. Their oscillator coordinates are $y_\star\simeq0.634$ and $0.770$, compared with xenon values near $1.75$ to $1.78$. The xenon and argon responses are shown in \cref{figtargets}. The shift reflects both the target dependence of the many body amplitudes and the conversion from cancellation momentum to recoil energy.

For a nuclear cancellation, the relevant target dependent momentum $q_\star$ satisfies
\begin{equation}
\frac12\mathcal M_M^{(1)}(q_\star)-\mathcal M_{\Phi''}^{(1)}(q_\star)\simeq0,
\qquad
E_\star^{\rm nuc}=\frac{q_\star^2}{2m_A},
\label{eqnuclearfeature}
\end{equation}
where $q_\star$ is the momentum transfer at the amplitude cancellation and $E_\star^{\rm nuc}$ is the corresponding recoil energy, with the isotope weights and multipole sum understood. Equation~\cref{eqnuclearfeature} displays the two sources of target dependence explicitly: shell occupancies and multipole matrix elements determine $q_\star$, while the nuclear mass converts this momentum into a recoil energy.

Endothermic scattering is characterized by a different scale. Minimizing \cref{eqvinel} with respect to the recoil energy gives
\begin{equation}
E_\star^{\rm endo}=\delta\frac{\mu_{\chi A}}{m_A},
\qquad
v_{\min,\star}^{\rm endo}=\sqrt{\frac{2\delta}{\mu_{\chi A}}}.
\label{eqendofeature}
\end{equation}
For heavy dark matter, $E_\star^{\rm endo}$ approaches the splitting $\delta$. A fixed excitation energy therefore gives increasingly similar characteristic recoil energies for different heavy targets. For $m_\chi=1~\mathrm{TeV}$ and $\delta=350~\mathrm{keV}$, the characteristic energies are about $312~\mathrm{keV}$ in $^{131}$Xe and $337~\mathrm{keV}$ in $^{40}$Ar. By comparison, the nuclear minima shift from $211$--$214~\mathrm{keV}$ in xenon to approximately $350$--$425~\mathrm{keV}$ in argon.

For targets $A$ and $B$ with nuclear masses $m_A$ and $m_B$, define the dimensionless endothermic target ratio $\mathcal T_{A/B}^{\rm endo}$. The splitting cancels from the ratio,
\begin{equation}
\mathcal T_{A/B}^{\rm endo}
=\frac{E_{\star,A}^{\rm endo}}{E_{\star,B}^{\rm endo}}
=\frac{m_\chi+m_B}{m_\chi+m_A}
\longrightarrow1
\quad (m_\chi\gg m_A,m_B).
\label{eqtargetratio}
\end{equation}
At $1~\mathrm{TeV}$, \cref{eqtargetratio} gives $\mathcal T_{\rm Ar/Xe}^{\rm endo}=1.082$. Pairing the two argon Hamiltonians with the GCN and JJ55 xenon nodes gives nuclear ratios from $1.63$ to $2.01$. A two target measurement can therefore test the reduced mass scaling expected for endothermic excitation before a splitting is inferred.

\begin{figure}[t]
\centering
\includegraphics[width=\columnwidth]{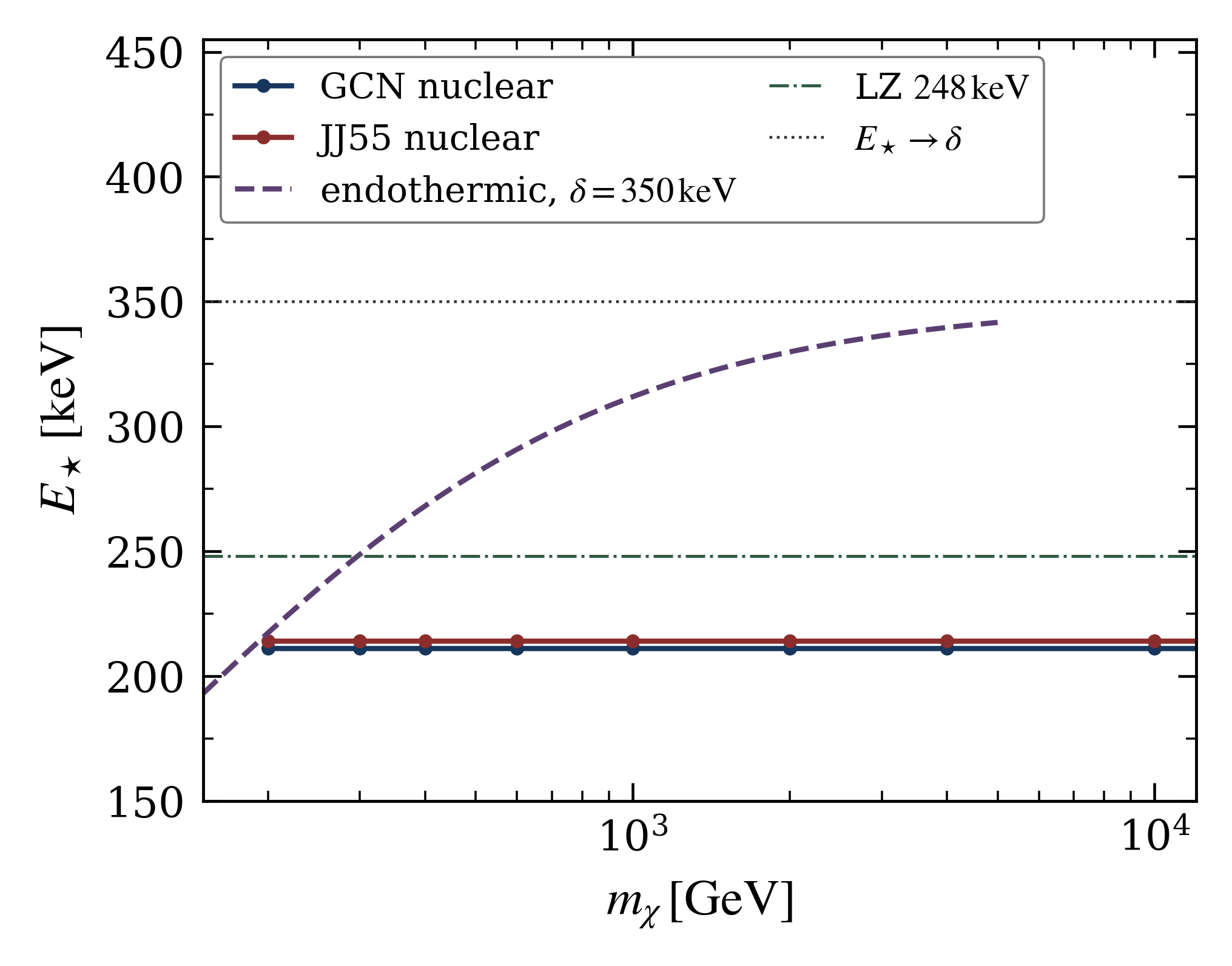}
\caption{Characteristic recoil scale versus dark matter mass. GCN and JJ55 nuclear minima remain essentially mass invariant. The endothermic scale for $\delta=350~\mathrm{keV}$ follows $\delta\mu_{\chi A}/m_A$ and approaches $\delta$ in the heavy limit. The horizontal line marks the reported LZ recoil energy~\cite{LZ2026highE}.}
\label{figscaledisc}
\end{figure}

The mass dependence in \cref{figscaledisc} separates the mechanisms without absolute rate normalization. Varying $m_\chi$ moves the endothermic characteristic recoil through the reduced mass, whereas the nuclear node remains tied to the xenon amplitude zero. Mass and target scaling therefore test separate consequences of assigning the recoil energy to an excitation threshold or a nuclear cancellation.

\section{Timing and supersymmetry inference}
\label{sectime}

Annual modulation probes the incident speed distribution at fixed recoil energy and target. We use the Standard Halo Model, with the truncated Maxwellian distribution and parameters listed in the Appendix. For $E_R=248~\mathrm{keV}$ and $m_\chi=1~\mathrm{TeV}$, elastic xenon scattering requires $v_{\min}^{\rm el}=339~\mathrm{km\,s^{-1}}$ from \cref{eqvmin}, which lies well inside the adopted speed distribution. Following the LZ halo convention~\cite{Baxter2021pqo}, we define $\mathcal A_{248}=(R_{\max}-R_{\min})/(R_{\max}+R_{\min})$, where $R_{\max}$ and $R_{\min}$ are the maximum and minimum annual rates. For the elastic interaction we obtain $\mathcal A_{248}=4.34\%$. At fixed recoil energy, the nuclear response multiplies the time dependent halo integral, so the fractional modulation is governed mainly by the incident speed distribution and depends only weakly on the GCN or JJ55 choice.

For endothermic scattering, the splitting $\delta$ raises $v_{\min}$. Splittings of $300$, $316$, and $350~\mathrm{keV}$ require approximately $705$, $724$, and $766~\mathrm{km\,s^{-1}}$ for the same $248~\mathrm{keV}$ recoil. The corresponding modulation fractions are about $34\%$, $42\%$, and $79\%$. With the adopted $^{131}$Xe mass convention, the kinematic endpoint lies near $\delta\simeq387~\mathrm{keV}$. Near this endpoint only the fastest part of the laboratory distribution contributes. Dedicated endothermic calculations find the associated large seasonal variation in this region~\cite{McCabe2026seasonal}.

\begin{figure}[t]
\centering
\includegraphics[width=\columnwidth]{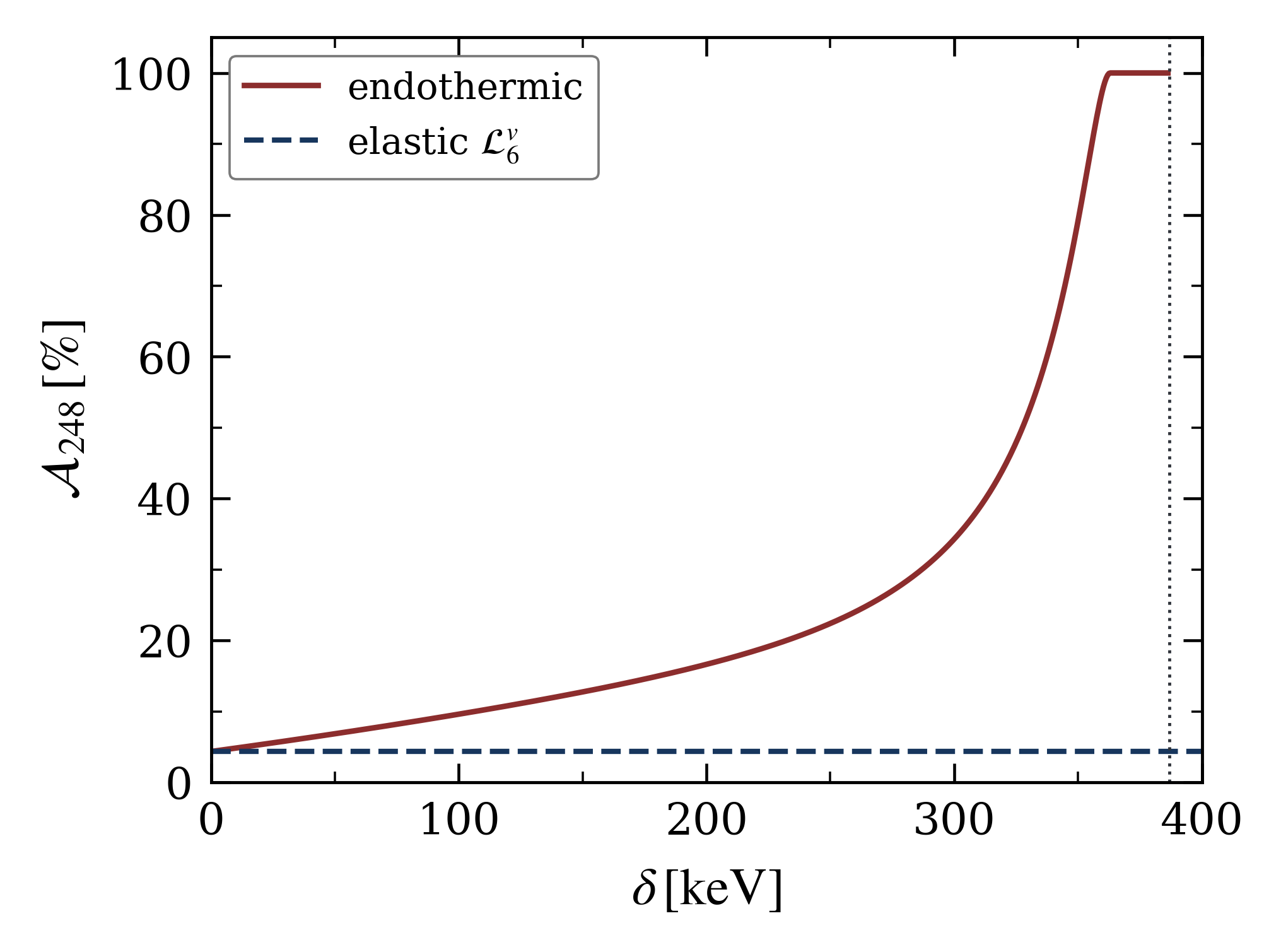}
\caption{Annual modulation amplitude at $248~\mathrm{keV}$ for elastic and endothermic scattering at $m_\chi=1~\mathrm{TeV}$. Elastic $\mathcal L_6^v$ remains weakly modulated. Endothermic scattering develops large time dependence as the dark state splitting drives the required incident speed toward the maximum laboratory halo speed.}
\label{figmod}
\end{figure}

Annual modulation is a standard probe of the local dark matter speed distribution~\cite{Drukier1986mod,Freese1988mod}. Stream calculations show how velocity substructure changes the seasonal recoil pattern~\cite{Savage2006streams}, while broader studies quantify astrophysical effects on direct detection kinematics~\cite{Freese2013modulation}. In the isotropic Standard Halo Model used here, the rate maximum remains close to the annual maximum of the laboratory speed for the elastic and endothermic points in \cref{figmod}. Changes in the local velocity distribution can alter modulation amplitude and phase, particularly near a kinematic cutoff~\cite{McCabe2010astro,Frandsen2012astro,Green2017astro}. The modulation should therefore be combined with the spectral position and target shift rather than used alone.

The timing comparison is also relevant to Higgsino interpretations. Heavy particle effective theory and electroweak loop calculations describe the weak scale, hadronic matching, and direct detection amplitudes of Wino and Higgsino states~\cite{HillSolon2014heavy,HillSolon2015one,HillSolon2015two}~\cite{Hisano2011ew,Hisano2013susy,NagataShirai2015}, with further matching to nucleon observables in Refs.~\cite{HisanoNagai2015,Bishara2020rge}. For a Higgsino, the neutral splitting arises from mixing with heavier neutralinos and constrains combinations of bino and wino masses together with supersymmetric boundary conditions. Recent analyses associate the event with a Higgsino mass near $1.1~\mathrm{TeV}$ and a splitting of a few hundred keV~\cite{FanReece2026lz,Freese2026lz,WuZhangZhu2026lz}, with inelastic strength set by the off diagonal electroweak interaction~\cite{DiMauro2026lz}. Generic mixing can place gauginos far above the Higgsino scale~\cite{FanReece2026lz,Yin2026lz,NagataShirai2015}, while a nonuniversal $SU(5)$ relation allows a wino near $120~\mathrm{TeV}$ for a $1.091~\mathrm{TeV}$ Higgsino~\cite{DuWang2026lz}. Peccei Quinn charged doublets and broader electroweak multiplets provide further inelastic realizations~\cite{Visinelli2026lz,Smirnov2026lz,Hisano2011ew}, and solar capture gives an independent constraint under the assumptions of Ref.~\cite{PospelovRamani2026lz}. Mapping a splitting onto a supersymmetric spectrum therefore requires the electroweak mixing pattern and ultraviolet boundary conditions~\cite{DuWang2026lz,Khan2025ibo,Muhammad2026gmg}.

In the elastic $\Lsix$ interpretation, no neutral state splitting is required to set the recoil scale; it is associated instead with the target dependent nuclear response. Large annual modulation would indicate recoils drawn mainly from the high speed tail~\cite{McCabe2026seasonal}. The high energy sideband gives an additional comparison. Thermal Higgsino fits to the event can produce recoils at larger reconstructed energy~\cite{Rodd2026sideband}. The $\Lsix$ spectrum also rises above its second node, but its energy dependence is fixed by the nuclear response. An acceptance corrected sideband shape can therefore compare the two spectral mechanisms.

\section{Interpretation and experimental tests}
\label{secdiscussion}

The calculation gives three observables probing different parts of the scattering process. The two xenon cancellations and the rise above the second node test the momentum dependence of the nuclear amplitude. The target shift compares many body matrix elements with reduced mass kinematics, while annual modulation tests the incident speed distribution. Together, these observables compare nuclear structure with dark sector excitation without using the overall event normalization.

The interpretations differ in the quantity setting the recoil energy. Endothermic and Higgsino scenarios associate the characteristic scale with an excitation threshold~\cite{DiMauro2026lz,Su2026lz,Yamashita2026lz}~\cite{FanReece2026lz,Freese2026lz,WuZhangZhu2026lz}, and electroweak multiplets provide related inelastic realizations~\cite{Smirnov2026lz}. Atmospheric neutrino upscattering introduces a kinematic cutoff~\cite{Jeesun2026nu}, fermionic absorption uses the dark matter rest energy~\cite{Lou2026lz}, and elastic pseudoscalar scattering through $\mathcal L_4$ probes a different momentum dependent nuclear response~\cite{Unwin2026lz}. For $\mathcal L_6^v$, the recoil structure follows instead from two cancellations between density and orbital spin orbit amplitudes, which can be related to isotope composition, target dependence, and annual modulation.

Combining the xenon spectrum, a second target, and annual modulation compares these scaling laws. In the heavy mass regime, endothermic scattering gives a target ratio near unity and large modulation as the required speed approaches the maximum halo speed. Nuclear interference gives a larger target shift and few percent elastic modulation in the adopted halo model. Feature positions, normalized shapes, target ratios, and modulation fractions are independent of the overall factor $(d_6^v)^2$, allowing shape and scaling tests before an experiment level likelihood determines the absolute coupling from the rate.

EFT analyses by LUX, XENON1T, and LZ have extended xenon searches beyond the standard coherent recoil template~\cite{LUX2021eft,XENON2024eft,LZ2024nreft}. Global NREFT studies also show how operator sensitivity changes across combined experimental and neutrino constraints~\cite{Catena2014uqa,AvisKozar2025global}. In the covariant LZ basis, coefficient ratios are fixed before the nuclear response is evaluated~\cite{LZ2024covariant}. For $\mathcal L_6^v$, the relations in \cref{eqcoeff} select a definite density and orbital spin orbit interference. The cancellation energy is therefore determined by this correlated interaction together with the nuclear matrix elements.

\section{Conclusion}
\label{secconclusion}

The 2026 LUX-ZEPLIN high energy recoil result corresponds to momentum transfer near $0.25~\mathrm{GeV}$, where finite size xenon structure contributes to the scattering amplitude. For the isovector $\mathcal L_6^v$ interaction considered by LZ, covariant matching removes the longitudinal $\Sigma''$ response and fixes the leading scalar contribution to interference between density and orbital spin orbit amplitudes. With the public GCN and JJ55 matrix elements, the response contains two finite momentum cancellations. The second lies near $211$--$214~\mathrm{keV}$, placing the reported recoil on the rising spectrum above this minimum.

The nuclear interpretation can be tested through independent scaling laws. The high xenon node is essentially unchanged across the heavy dark matter regime, and the curvature weighted isotope centroid reproduces its natural xenon position and residual depth. Changing target gives nuclear Ar to Xe ratios from $1.63$ to $2.01$ for the displayed Hamiltonians, compared with $1.082$ for a $1~\mathrm{TeV}$ endothermic example. The difference reflects nuclear structure in one case and reduced mass scaling in the other.

The spectral shape and annual modulation add direct measurements. The xenon rate gives $R(270)/R(248)=1.73$ for GCN and $1.85$ for JJ55, while elastic $\mathcal L_6^v$ scattering has a modulation fraction near $4.34\%$ at $248~\mathrm{keV}$ in the adopted Standard Halo Model. Endothermic solutions close to the maximum halo speed show much larger seasonal variation. Additional xenon exposure can test the cancellation regions and high energy rise, a second target the target scaling, and time resolved data the required incident speeds. Together, these measurements can determine whether the recoil scale is associated with a finite momentum nuclear response or a dark sector excitation before mapping any neutral state splitting onto a supersymmetric spectrum.

\section*{Acknowledgements}
  SC  acknowledges the support of  Istituto Nazionale di Fisica Nucleare (INFN) Sezioni  di Napoli e di Frascati, {\it Iniziative Specifiche} QGSKY,  MOONLIGHT2 and  the Gruppo Nazionale di Fisica Matematica (GNFM)  of Istituto Nazionale di Alta Matematica (INDAM).

\section*{AI Declaration}
ChatGPT was used solely for language editing, while the authors take full responsibility for the scientific content and final manuscript.
\appendix
\section{Nuclear contraction and halo functions}
\label{appchecks}

The nuclear contraction uses the one body density matrices distributed with the response calculation and the harmonic oscillator single particle multipoles of the response formalism~\cite{Gorton2022cpc}. Isovector amplitudes use one half of the proton minus neutron density, and each isotope contribution contains the target spin average $1/(2J_A+1)$. With the response calculation mass convention, the second minima are $E_{\star,2}=211.07~\mathrm{keV}$ for GCN and $214.02~\mathrm{keV}$ for JJ55. The transverse spin fractions at these minima are $1.83\times10^{-4}$ and $5.07\times10^{-4}$.

\begin{figure}[t]
\centering
\includegraphics[width=\columnwidth]{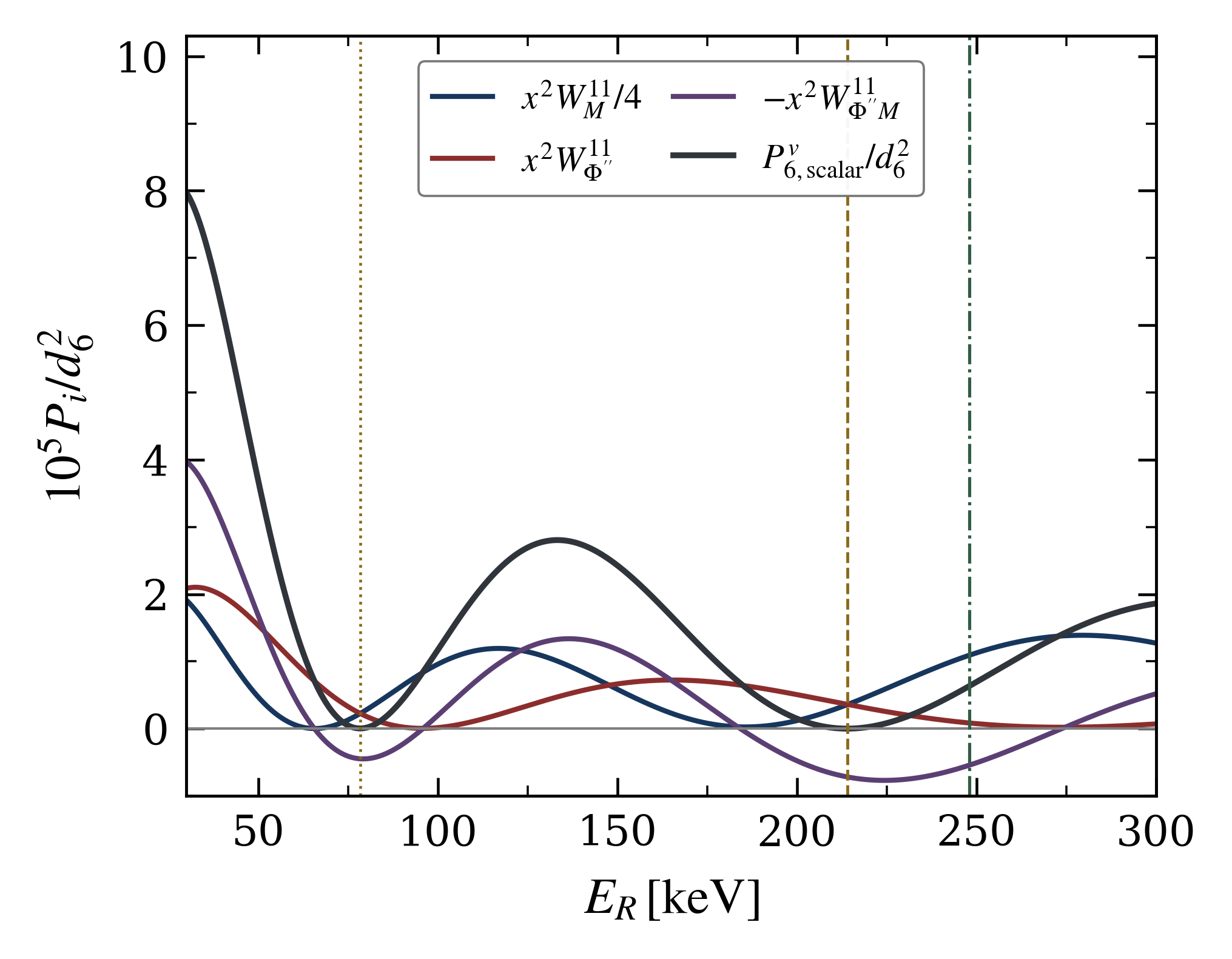}
\caption{JJ55 natural xenon response corresponding to \cref{figresponses}. Density, spin orbit, and interference contributions generate the second cancellation at $214~\mathrm{keV}$.}
\label{figjj55resp}
\end{figure}

The JJ55 decomposition in \cref{figjj55resp} has the same response structure as GCN, with the second cancellation shifted by about $3~\mathrm{keV}$.

For the truncated Maxwellian Standard Halo Model, $v_0$ denotes the characteristic halo speed, $v_{\rm esc}$ the Galactic escape speed, and $v_e(t)$ the time dependent laboratory speed. We define $z=v_{\rm esc}/v_0$ and the normalization $N_{\rm esc}$ by
\begin{equation}
z=\frac{v_{\rm esc}}{v_0},
\qquad
N_{\rm esc}=\operatorname{erf}(z)-\frac{2z}{\sqrt{\pi}}e^{-z^2}.
\label{eqnesc}
\end{equation}
Here $\operatorname{erf}$ is the error function. The differential recoil rate contains the mean inverse speed $\eta(v_{\min},v_e)$. For $v_{\min}<v_{\rm esc}-v_e$, this function is
\begin{align}
2N_{\rm esc}v_e\,\eta(v_{\min},v_e)={}&
\operatorname{erf}\!\left(\frac{v_{\min}+v_e}{v_0}\right)
-\operatorname{erf}\!\left(\frac{v_{\min}-v_e}{v_0}\right)\nonumber\\
&-\frac{4v_e}{\sqrt{\pi}v_0}e^{-z^2},
\label{eqeta1}
\end{align}
whereas for $v_{\rm esc}-v_e\le v_{\min}<v_{\rm esc}+v_e$ it becomes
\begin{align}
2N_{\rm esc}v_e\,\eta(v_{\min},v_e)={}&
\operatorname{erf}(z)-\operatorname{erf}\!\left(\frac{v_{\min}-v_e}{v_0}\right)\nonumber\\
&-\frac{2(v_{\rm esc}-v_{\min}+v_e)}{\sqrt{\pi}v_0}e^{-z^2}.
\label{eqeta2}
\end{align}
Equations~\cref{eqeta1,eqeta2} are the two kinematic branches of the mean inverse speed, and the rate vanishes for $v_{\min}>v_{\rm esc}+v_e$. We use $v_0=238~\mathrm{km\,s^{-1}}$ and $v_{\rm esc}=544~\mathrm{km\,s^{-1}}$, together with the solar and terrestrial velocities specified in the recommended direct detection convention~\cite{Baxter2021pqo,Lewin1995rx}.

The response basis and relativistic matching follow direct detection EFT constructions~\cite{DelNobile2018dfm,Bishara2016hnh,Brod2017bsw}. Weak scale and quantum chromodynamics (QCD) matching relate electroweak interactions to hadronic matrix elements~\cite{HillSolon2015one,HillSolon2015two,HisanoNagai2015}, while renormalization group evolution connects scales for more general dark matter interactions~\cite{Bishara2020rge}. The nuclear calculation retains the $M$, $\Phi''$, $\Phi''M$, and $\Sigma'$ contributions generated by \cref{eqcoeff}. Global NREFT analyses illustrate the target dependence of noncoherent channels~\cite{Catena2014uqa}, while chiral and many body calculations extend the response analysis across interaction classes~\cite{Hoferichter2015chiral,Hoferichter2016analysis,AbdelKhaleq2025cevns}.

\bibliography{references}

@misc{LZ2026highE,
  author = {{LZ Collaboration}},
  title = {Search for dark matter particle interactions in an extended nuclear recoil energy window with the LUX-ZEPLIN (LZ) experiment},
  eprint = {2609.02823},
  archivePrefix = {arXiv},
  primaryClass = {hep-ex},
  year = {2026},
  month = {9}
}

@article{LZ2024covariant,
  author = {{LZ Collaboration}},
  title = {Constraints on Covariant Dark-Matter--Nucleon Effective Field Theory Interactions from the First Science Run of the LUX-ZEPLIN Experiment},
  journal = {Phys. Rev. Lett.},
  volume = {133},
  pages = {221801},
  year = {2024},
  doi = {10.1103/PhysRevLett.133.221801}
}

@article{LZ2024nreft,
  author = {{LZ Collaboration}},
  title = {First constraints on WIMP-nucleon effective field theory couplings in an extended energy region from LUX-ZEPLIN},
  journal = {Phys. Rev. D},
  volume = {109},
  pages = {092003},
  year = {2024},
  eprint = {2312.02030},
  archivePrefix = {arXiv},
  primaryClass = {hep-ex},
  doi = {10.1103/PhysRevD.109.092003}
}

@article{Fitzpatrick2012ix,
  author = {Fitzpatrick, A. Liam and Haxton, Wick and Katz, Emanuel and Lubbers, Nicholas and Xu, Yiming},
  title = {The Effective Field Theory of Dark Matter Direct Detection},
  journal = {JCAP},
  volume = {02},
  number = {02},
  pages = {004},
  year = {2013},
  eprint = {1203.3542},
  archivePrefix = {arXiv},
  primaryClass = {hep-ph},
  doi = {10.1088/1475-7516/2013/02/004}
}

@article{Anand2013yka,
  author = {Anand, Nikhil and Fitzpatrick, A. Liam and Haxton, W. C.},
  title = {Weakly interacting massive particle-nucleus elastic scattering response},
  journal = {Phys. Rev. C},
  volume = {89},
  pages = {065501},
  year = {2014},
  eprint = {1308.6288},
  archivePrefix = {arXiv},
  primaryClass = {hep-ph},
  doi = {10.1103/PhysRevC.89.065501}
}

@article{DelNobile2018dfm,
  author = {Del Nobile, Eugenio},
  title = {Complete Lorentz-to-Galileo dictionary for direct dark matter detection},
  journal = {Phys. Rev. D},
  volume = {98},
  pages = {123003},
  year = {2018},
  eprint = {1806.01291},
  archivePrefix = {arXiv},
  primaryClass = {hep-ph},
  doi = {10.1103/PhysRevD.98.123003}
}

@article{Gresham2014vja,
  author = {Gresham, Moira I. and Zurek, Kathryn M.},
  title = {On the Effect of Nuclear Response Functions in Dark Matter Direct Detection},
  journal = {Phys. Rev. D},
  volume = {89},
  pages = {123521},
  year = {2014},
  eprint = {1401.3739},
  archivePrefix = {arXiv},
  primaryClass = {hep-ph},
  doi = {10.1103/PhysRevD.89.123521}
}

@article{Catena2014uqa,
  author = {Catena, Riccardo and Gondolo, Paolo},
  title = {Global fits of the dark matter-nucleon effective interactions},
  journal = {JCAP},
  volume = {09},
  number = {09},
  pages = {045},
  year = {2014},
  eprint = {1405.2637},
  archivePrefix = {arXiv},
  primaryClass = {hep-ph},
  doi = {10.1088/1475-7516/2014/09/045}
}

@article{Gorton2022cpc,
  author = {Gorton, Oliver C. and Johnson, Calvin W. and Jiao, Changfeng and Nikoleyczik, Jonathan},
  title = {dmscatter: A fast program for WIMP-nucleus scattering},
  journal = {Comput. Phys. Commun.},
  volume = {284},
  pages = {108597},
  year = {2023},
  eprint = {2209.09187},
  archivePrefix = {arXiv},
  primaryClass = {nucl-th},
  doi = {10.1016/j.cpc.2022.108597}
}

@article{Vietze2014jga,
  author = {Vietze, Laura and Klos, Philipp and Menendez, Javier and Haxton, Wick C. and Schwenk, Achim},
  title = {Nuclear structure aspects of spin-independent WIMP scattering off xenon},
  journal = {Phys. Rev. D},
  volume = {91},
  pages = {043520},
  year = {2015},
  eprint = {1412.6091},
  archivePrefix = {arXiv},
  primaryClass = {nucl-th},
  doi = {10.1103/PhysRevD.91.043520}
}

@article{AbdelKhaleq2024nuc,
  author = {Abdel Khaleq, R. and Busoni, G. and Simenel, C. and Stuchbery, A. E.},
  title = {Impact of shell model interactions on nuclear responses to WIMP elastic scattering},
  journal = {Phys. Rev. D},
  volume = {109},
  pages = {075036},
  year = {2024},
  eprint = {2311.15764},
  archivePrefix = {arXiv},
  primaryClass = {hep-ph},
  doi = {10.1103/PhysRevD.109.075036}
}

@article{Baxter2021pqo,
  author = {Baxter, D. and Bloch, I. M. and Bodnia, E. and others},
  title = {Recommended conventions for reporting results from direct dark matter searches},
  journal = {Eur. Phys. J. C},
  volume = {81},
  pages = {907},
  year = {2021},
  eprint = {2105.00599},
  archivePrefix = {arXiv},
  primaryClass = {hep-ex},
  doi = {10.1140/epjc/s10052-021-09655-y}
}

@article{TuckerSmith2001hy,
  author = {Tucker-Smith, David and Weiner, Neal},
  title = {Inelastic dark matter},
  journal = {Phys. Rev. D},
  volume = {64},
  pages = {043502},
  year = {2001},
  eprint = {hep-ph/0101138},
  doi = {10.1103/PhysRevD.64.043502}
}

@article{TuckerSmith2004jv,
  author = {Tucker-Smith, David and Weiner, Neal},
  title = {Status of inelastic dark matter},
  journal = {Phys. Rev. D},
  volume = {72},
  pages = {063509},
  year = {2005},
  eprint = {hep-ph/0402065},
  doi = {10.1103/PhysRevD.72.063509}
}

@article{Barello2014eaa,
  author = {Barello, G. and Chang, Spencer and Newby, Christopher A.},
  title = {A Model Independent Approach to Inelastic Dark Matter Scattering},
  journal = {Phys. Rev. D},
  volume = {90},
  pages = {094027},
  year = {2014},
  eprint = {1409.0536},
  archivePrefix = {arXiv},
  primaryClass = {hep-ph},
  doi = {10.1103/PhysRevD.90.094027}
}

@article{Bramante2016rdh,
  author = {Bramante, Joseph and Fox, Patrick J. and Kribs, Graham D. and Martin, Adam},
  title = {Inelastic frontier: Discovering dark matter at high recoil energy},
  journal = {Phys. Rev. D},
  volume = {94},
  pages = {115026},
  year = {2016},
  eprint = {1608.02662},
  archivePrefix = {arXiv},
  primaryClass = {hep-ph},
  doi = {10.1103/PhysRevD.94.115026}
}

@misc{DiMauro2026lz,
  author = {Di Mauro, Mattia},
  title = {Dark Matter at the Kinematic Edge: Interpreting the 248 keV LZ Nuclear-Recoil Candidate},
  eprint = {2609.02608},
  archivePrefix = {arXiv},
  primaryClass = {hep-ph},
  year = {2026},
  month = {9}
}

@misc{Freese2026lz,
  author = {Freese, Katherine and Theodosopoulos, Dionysios P.},
  title = {Higgsino Dark Matter Interpretation of the LUX-ZEPLIN 248 keV Nuclear-Recoil Event},
  eprint = {2609.01583},
  archivePrefix = {arXiv},
  primaryClass = {hep-ph},
  year = {2026},
  month = {9}
}

@misc{Yin2026lz,
  author = {Yin, Wen},
  title = {A PQ-Symmetric High-Scale SUSY Interpretation of the LZ High-Energy Recoil},
  eprint = {2609.01892},
  archivePrefix = {arXiv},
  primaryClass = {hep-ph},
  year = {2026},
  month = {9}
}

@misc{DuWang2026lz,
  author = {Du, Xiaokang and Wang, Fei},
  title = {TeV Higgsino Interpretation of the LZ High-Recoil Event with Intermediate-Scale Electroweak Gauginos},
  eprint = {2609.04163},
  archivePrefix = {arXiv},
  primaryClass = {hep-ph},
  year = {2026},
  month = {9}
}

@misc{Su2026lz,
  author = {Su, Liangliang and Yang, Jin Min and Yang, Wen-Na},
  title = {Inelastic Dark Matter Signature at High Recoil Energy in LUX-ZEPLIN and CRESST},
  eprint = {2609.01475},
  archivePrefix = {arXiv},
  primaryClass = {hep-ph},
  year = {2026},
  month = {9}
}

@misc{Lou2026lz,
  author = {Lou, Yuanchao and Lu, Chih-Ting},
  title = {Fermionic Dark Matter Absorption and the High-Energy Event in LUX-ZEPLIN},
  eprint = {2609.01592},
  archivePrefix = {arXiv},
  primaryClass = {hep-ph},
  year = {2026},
  month = {9}
}

@misc{McCabe2026seasonal,
  author = {McCabe, Christopher},
  title = {Seasonal dark matter from the LUX-ZEPLIN high-energy event},
  eprint = {2609.04181},
  archivePrefix = {arXiv},
  primaryClass = {hep-ph},
  year = {2026},
  month = {9}
}

@article{Fan2010gt,
  author = {Fan, JiJi and Reece, Matthew and Wang, Lian-Tao},
  title = {Non-relativistic effective theory of dark matter direct detection},
  journal = {JCAP},
  volume = {11},
  number = {11},
  pages = {042},
  year = {2010},
  eprint = {1008.1591},
  archivePrefix = {arXiv},
  primaryClass = {hep-ph},
  doi = {10.1088/1475-7516/2010/11/042}
}

@misc{Fitzpatrick2012ib,
  author = {Fitzpatrick, A. Liam and Haxton, Wick and Katz, Emanuel and Lubbers, Nicholas and Xu, Yiming},
  title = {Model Independent Direct Detection Analyses},
  year = {2012},
  eprint = {1211.2818},
  archivePrefix = {arXiv},
  primaryClass = {hep-ph}
}

@article{Bishara2016hnh,
  author = {Bishara, Fady and Brod, Joachim and Grinstein, Benjamin and Zupan, Jure},
  title = {Chiral Effective Theory of Dark Matter Direct Detection},
  journal = {JCAP},
  volume = {02},
  number = {02},
  pages = {009},
  year = {2017},
  eprint = {1611.00368},
  archivePrefix = {arXiv},
  primaryClass = {hep-ph},
  doi = {10.1088/1475-7516/2017/02/009}
}

@article{Brod2017bsw,
  author = {Brod, Joachim and Gootjes-Dreesbach, Aaron and Tammaro, Michele and Zupan, Jure},
  title = {Effective Field Theory for Dark Matter Direct Detection up to Dimension Seven},
  journal = {JHEP},
  volume = {10},
  number = {10},
  pages = {065},
  year = {2018},
  eprint = {1710.10218},
  archivePrefix = {arXiv},
  primaryClass = {hep-ph},
  doi = {10.1007/JHEP10(2018)065}
}

@article{Menendez2012tm,
  author = {Menendez, Javier and Gazit, Doron and Schwenk, Achim},
  title = {Spin-dependent WIMP scattering off nuclei},
  journal = {Phys. Rev. D},
  volume = {86},
  pages = {103511},
  year = {2012},
  eprint = {1208.1094},
  archivePrefix = {arXiv},
  primaryClass = {astro-ph.CO},
  doi = {10.1103/PhysRevD.86.103511}
}

@article{Klos2013rwa,
  author = {Klos, Philipp and Menendez, Javier and Gazit, Doron and Schwenk, Achim},
  title = {Large-scale nuclear structure calculations for spin-dependent WIMP scattering with chiral effective field theory currents},
  journal = {Phys. Rev. D},
  volume = {88},
  pages = {083516},
  year = {2013},
  eprint = {1304.7684},
  archivePrefix = {arXiv},
  primaryClass = {nucl-th},
  doi = {10.1103/PhysRevD.88.083516}
}

@article{Hoferichter2018acd,
  author = {Hoferichter, Martin and Klos, Philipp and Menendez, Javier and Schwenk, Achim},
  title = {Nuclear structure factors for general spin-independent WIMP-nucleus scattering},
  journal = {Phys. Rev. D},
  volume = {99},
  pages = {055031},
  year = {2019},
  eprint = {1812.05617},
  archivePrefix = {arXiv},
  primaryClass = {hep-ph},
  doi = {10.1103/PhysRevD.99.055031}
}

@article{Lewin1995rx,
  author = {Lewin, J. D. and Smith, P. F.},
  title = {Review of mathematics, numerical factors, and corrections for dark matter experiments based on elastic nuclear recoil},
  journal = {Astropart. Phys.},
  volume = {6},
  pages = {87--112},
  year = {1996},
  doi = {10.1016/S0927-6505(96)00047-3}
}

@misc{Unwin2026lz,
  author = {Unwin, James},
  title = {Axion Portal Dark Matter and the LUX-ZEPLIN High-Recoil Event},
  year = {2026},
  month = {9},
  eprint = {2609.04186},
  archivePrefix = {arXiv},
  primaryClass = {hep-ph}
}

@article{Bozorgnia2018dvr,
  author = {Bozorgnia, Nassim and Cerdeno, David G. and Cheek, Andrew and Penning, Bjoern},
  title = {Opening the energy window on direct dark matter detection},
  journal = {JCAP},
  volume = {12},
  number = {12},
  pages = {013},
  year = {2018},
  eprint = {1810.05576},
  archivePrefix = {arXiv},
  primaryClass = {hep-ph},
  doi = {10.1088/1475-7516/2018/12/013}
}

@article{Freese2013modulation,
  author = {Freese, Katherine and Lisanti, Mariangela and Savage, Christopher},
  title = {Colloquium: Annual modulation of dark matter},
  journal = {Rev. Mod. Phys.},
  volume = {85},
  pages = {1561--1581},
  year = {2013},
  eprint = {1209.3339},
  archivePrefix = {arXiv},
  primaryClass = {astro-ph.CO},
  doi = {10.1103/RevModPhys.85.1561}
}

@article{AbdelKhaleq2022natXe,
  author = {Abdel Khaleq, Raghda and Simenel, Cedric and Stuchbery, Andrew E.},
  title = {Impact of nuclear structure from shell model calculations on nuclear responses to WIMP elastic scattering for $^{19}$F and $^{nat}$Xe targets},
  journal = {SciPost Phys. Proc.},
  volume = {12},
  pages = {062},
  year = {2023},
  eprint = {2209.15250},
  archivePrefix = {arXiv},
  primaryClass = {hep-ph},
  doi = {10.21468/SciPostPhysProc.12.062}
}

@article{AbdelKhaleq2025cevns,
  author = {Abdel Khaleq, Raghda and Newstead, Jayden L. and Simenel, Cedric and Stuchbery, Andrew E.},
  title = {Detailed nuclear structure calculations for coherent elastic neutrino-nucleus scattering},
  journal = {Phys. Rev. D},
  volume = {111},
  pages = {033003},
  year = {2025},
  eprint = {2405.20060},
  archivePrefix = {arXiv},
  primaryClass = {hep-ph},
  doi = {10.1103/PhysRevD.111.033003}
}

@misc{Rodd2026sideband,
  author = {Rodd, Nicholas L. and Safdi, Benjamin R. and Slatyer, Tracy R. and Xu, Weishuang Linda},
  title = {Confronting the Higgsino Interpretation of the LZ Event with the High-Energy Sideband},
  eprint = {2609.04175},
  archivePrefix = {arXiv},
  primaryClass = {hep-ph},
  year = {2026},
  month = {9}
}

@misc{Jeesun2026nu,
  author = {Jeesun, Sk and Majumdar, Anirban},
  title = {Atmospheric neutrino up-scattering explanation of LZ 2026 excess},
  eprint = {2609.04185},
  archivePrefix = {arXiv},
  primaryClass = {hep-ph},
  year = {2026},
  month = {9}
}

@misc{FanReece2026lz,
  author = {Fan, JiJi and Reece, Matthew},
  title = {Higgsino Above the Sea of Fog},
  eprint = {2609.01504},
  archivePrefix = {arXiv},
  primaryClass = {hep-ph},
  year = {2026},
  month = {9}
}

@misc{WuZhangZhu2026lz,
  author = {Wu, Lei and Zhang, Yang and Zhu, Bin},
  title = {TeV Higgsino Dark Matter from LZ Nuclear Recoil to Fermi-LAT Gamma Rays},
  eprint = {2609.01590},
  archivePrefix = {arXiv},
  primaryClass = {hep-ph},
  year = {2026},
  month = {9}
}

@misc{PospelovRamani2026lz,
  author = {Pospelov, Maxim and Ramani, Harikrishnan},
  title = {Strong Constraints on Higgsino Dark Matter from Solar Capture},
  eprint = {2609.02775},
  archivePrefix = {arXiv},
  primaryClass = {hep-ph},
  year = {2026},
  month = {9}
}

@misc{Visinelli2026lz,
  author = {Visinelli, Luca},
  title = {A Peccei--Quinn Origin for Inelastic Electroweak Dark Matter after LUX-ZEPLIN},
  eprint = {2609.02807},
  archivePrefix = {arXiv},
  primaryClass = {hep-ph},
  year = {2026},
  month = {9}
}

@misc{Yamashita2026lz,
  author = {Yamashita, Kimiko},
  title = {Inelastic Dark Photon Dark Matter for the LUX-ZEPLIN High-Recoil Event and the Galactic Halo Gamma-Ray Excess},
  eprint = {2609.02868},
  archivePrefix = {arXiv},
  primaryClass = {hep-ph},
  year = {2026},
  month = {9}
}

@misc{Smirnov2026lz,
  author = {Smirnov, Juri and Griffith, Spencer and Beacom, John F.},
  title = {Inelastic Signatures of Electroweak Dark Matter},
  eprint = {2609.04144},
  archivePrefix = {arXiv},
  primaryClass = {hep-ph},
  year = {2026},
  month = {9}
}

@article{Khan2025ibo,
    author = "Khan, Imtiaz and Muhammad, Ali and Li, Tianjun and Raza, Shabbar and Pirzada and Khan, Mussawir",
    title = "{The light neutralino dark matter at Future Colliders in the MSSM with the generalized minimal supergravity (GmSUGRA)}",
    eprint = "2509.23356",
    archivePrefix = "arXiv",
    primaryClass = "hep-ph",
    doi = "10.1007/JHEP06(2026)115",
    journal = "JHEP",
    volume = "06",
    pages = "115",
    year = "2026"
}

@article{Muhammad2026gmg,
    author = "Muhammad, Ali and Khan, Imtiaz and Li, Tianjun and Raza, Shabbar and Khan, Mussawir and Pirzada",
    title = "{LHC Run-3, dark matter, and supersymmetric spectra in the supersymmetric Pati-Salam model}",
    eprint = "2603.24152",
    archivePrefix = "arXiv",
    primaryClass = "hep-ph",
    doi = "10.1016/j.physletb.2026.140890",
    journal = "Phys. Lett. B",
    volume = "881",
    pages = "140890",
    year = "2026"
}

@article{Chang2009idm,
  author = {Chang, Spencer and Kribs, Graham D. and Tucker-Smith, David and Weiner, Neal},
  title = {Inelastic Dark Matter in Light of DAMA/LIBRA},
  journal = {Phys. Rev. D},
  volume = {79},
  pages = {043513},
  year = {2009},
  eprint = {0807.2250},
  archivePrefix = {arXiv},
  primaryClass = {hep-ph},
  doi = {10.1103/PhysRevD.79.043513}
}

@article{Hoferichter2015chiral,
  author = {Hoferichter, Martin and Klos, Philipp and Schwenk, Achim},
  title = {Chiral power counting of one- and two-body currents in direct detection of dark matter},
  journal = {Phys. Lett. B},
  volume = {746},
  pages = {410--416},
  year = {2015},
  eprint = {1503.04811},
  archivePrefix = {arXiv},
  primaryClass = {hep-ph},
  doi = {10.1016/j.physletb.2015.05.041}
}

@article{Hoferichter2016analysis,
  author = {Hoferichter, Martin and Klos, Philipp and Menendez, Javier and Schwenk, Achim},
  title = {Analysis strategies for general spin-independent WIMP-nucleus scattering},
  journal = {Phys. Rev. D},
  volume = {94},
  pages = {063505},
  year = {2016},
  eprint = {1605.08043},
  archivePrefix = {arXiv},
  primaryClass = {hep-ph},
  doi = {10.1103/PhysRevD.94.063505}
}

@article{LUX2021eft,
  author = {Akerib, D. S. and others},
  collaboration = {LUX},
  title = {Effective field theory analysis of the first LUX dark matter search},
  journal = {Phys. Rev. D},
  volume = {103},
  pages = {122005},
  year = {2021},
  eprint = {2003.11141},
  archivePrefix = {arXiv},
  primaryClass = {astro-ph.CO},
  doi = {10.1103/PhysRevD.103.122005}
}

@article{XENON2024eft,
  author = {Aprile, E. and others},
  collaboration = {XENON},
  title = {Effective field theory and inelastic dark matter results from XENON1T},
  journal = {Phys. Rev. D},
  volume = {109},
  pages = {112017},
  year = {2024},
  eprint = {2210.07591},
  archivePrefix = {arXiv},
  primaryClass = {hep-ex},
  doi = {10.1103/PhysRevD.109.112017}
}

@article{Drukier1986mod,
  author = {Drukier, A. K. and Freese, Katherine and Spergel, D. N.},
  title = {Detecting Cold Dark Matter Candidates},
  journal = {Phys. Rev. D},
  volume = {33},
  pages = {3495--3508},
  year = {1986},
  doi = {10.1103/PhysRevD.33.3495}
}

@article{Freese1988mod,
  author = {Freese, Katherine and Frieman, Joshua A. and Gould, Andrew},
  title = {Signal Modulation in Cold Dark Matter Detection},
  journal = {Phys. Rev. D},
  volume = {37},
  pages = {3388--3405},
  year = {1988},
  doi = {10.1103/PhysRevD.37.3388}
}

@article{Savage2006streams,
  author = {Savage, Christopher and Freese, Katherine and Gondolo, Paolo},
  title = {Annual Modulation of Dark Matter in the Presence of Streams},
  journal = {Phys. Rev. D},
  volume = {74},
  pages = {043531},
  year = {2006},
  eprint = {astro-ph/0607121},
  archivePrefix = {arXiv},
  doi = {10.1103/PhysRevD.74.043531}
}

@article{McCabe2010astro,
  author = {McCabe, Christopher},
  title = {The Astrophysical Uncertainties Of Dark Matter Direct Detection Experiments},
  journal = {Phys. Rev. D},
  volume = {82},
  pages = {023530},
  year = {2010},
  eprint = {1005.0579},
  archivePrefix = {arXiv},
  primaryClass = {hep-ph},
  doi = {10.1103/PhysRevD.82.023530}
}

@article{Frandsen2012astro,
  author = {Frandsen, Mads T. and Kahlhoefer, Felix and McCabe, Christopher and Sarkar, Subir and Schmidt-Hoberg, Kai},
  title = {Resolving astrophysical uncertainties in dark matter direct detection},
  journal = {JCAP},
  volume = {01},
  pages = {024},
  year = {2012},
  eprint = {1111.0292},
  archivePrefix = {arXiv},
  primaryClass = {hep-ph},
  doi = {10.1088/1475-7516/2012/01/024}
}

@article{Green2017astro,
  author = {Green, Anne M.},
  title = {Astrophysical uncertainties on the local dark matter distribution and direct detection experiments},
  journal = {J. Phys. G},
  volume = {44},
  pages = {084001},
  year = {2017},
  eprint = {1703.10102},
  archivePrefix = {arXiv},
  primaryClass = {astro-ph.CO},
  doi = {10.1088/1361-6471/aa7819}
}

@article{HillSolon2014heavy,
  author = {Hill, Richard J. and Solon, Mikhail P.},
  title = {WIMP-nucleon scattering with heavy WIMP effective theory},
  journal = {Phys. Rev. Lett.},
  volume = {112},
  pages = {211602},
  year = {2014},
  eprint = {1309.4092},
  archivePrefix = {arXiv},
  primaryClass = {hep-ph},
  doi = {10.1103/PhysRevLett.112.211602}
}

@article{HillSolon2015one,
  author = {Hill, Richard J. and Solon, Mikhail P.},
  title = {Standard Model anatomy of WIMP dark matter direct detection I. Weak-scale matching},
  journal = {Phys. Rev. D},
  volume = {91},
  pages = {043504},
  year = {2015},
  eprint = {1401.3339},
  archivePrefix = {arXiv},
  primaryClass = {hep-ph},
  doi = {10.1103/PhysRevD.91.043504}
}

@article{HillSolon2015two,
  author = {Hill, Richard J. and Solon, Mikhail P.},
  title = {Standard Model anatomy of WIMP dark matter direct detection II. QCD analysis and hadronic matrix elements},
  journal = {Phys. Rev. D},
  volume = {91},
  pages = {043505},
  year = {2015},
  eprint = {1409.8290},
  archivePrefix = {arXiv},
  primaryClass = {hep-ph},
  doi = {10.1103/PhysRevD.91.043505}
}

@article{Hisano2011ew,
  author = {Hisano, Junji and Ishiwata, Koji and Nagata, Natsumi and Takesako, Tomohiro},
  title = {Direct Detection of Electroweak-Interacting Dark Matter},
  journal = {JHEP},
  volume = {07},
  pages = {005},
  year = {2011},
  eprint = {1104.0228},
  archivePrefix = {arXiv},
  primaryClass = {hep-ph},
  doi = {10.1007/JHEP07(2011)005}
}

@article{Hisano2013susy,
  author = {Hisano, Junji and Ishiwata, Koji and Nagata, Natsumi},
  title = {Direct Search of Dark Matter in High-Scale Supersymmetry},
  journal = {Phys. Rev. D},
  volume = {87},
  pages = {035020},
  year = {2013},
  eprint = {1210.5985},
  archivePrefix = {arXiv},
  primaryClass = {hep-ph},
  doi = {10.1103/PhysRevD.87.035020}
}

@article{NagataShirai2015,
  author = {Nagata, Natsumi and Shirai, Satoshi},
  title = {Higgsino Dark Matter in High-Scale Supersymmetry},
  journal = {JHEP},
  volume = {01},
  pages = {029},
  year = {2015},
  eprint = {1410.4549},
  archivePrefix = {arXiv},
  primaryClass = {hep-ph},
  doi = {10.1007/JHEP01(2015)029}
}

@article{HisanoNagai2015,
  author = {Hisano, Junji and Nagai, Ryo and Nagata, Natsumi},
  title = {Effective Theories for Dark Matter Nucleon Scattering},
  journal = {JHEP},
  volume = {05},
  pages = {037},
  year = {2015},
  eprint = {1502.02244},
  archivePrefix = {arXiv},
  primaryClass = {hep-ph},
  doi = {10.1007/JHEP05(2015)037}
}

@article{Bishara2020rge,
  author = {Bishara, Fady and Brod, Joachim and Grinstein, Benjamin and Zupan, Jure},
  title = {Renormalization Group Effects in Dark Matter Interactions},
  journal = {JHEP},
  volume = {03},
  pages = {089},
  year = {2020},
  eprint = {1809.03506},
  archivePrefix = {arXiv},
  primaryClass = {hep-ph},
  doi = {10.1007/JHEP03(2020)089}
}

@article{Cirelli2013tools,
  author = {Cirelli, Marco and Del Nobile, Eugenio and Panci, Paolo},
  title = {Tools for model-independent bounds in direct dark matter searches},
  journal = {JCAP},
  volume = {10},
  pages = {019},
  year = {2013},
  eprint = {1307.5955},
  archivePrefix = {arXiv},
  primaryClass = {hep-ph},
  doi = {10.1088/1475-7516/2013/10/019}
}

@article{AvisKozar2025global,
  author = {Avis Kozar, Neal P. and Scott, Pat and Vincent, Aaron C.},
  title = {A global fit of non-relativistic effective dark matter operators including solar neutrinos},
  journal = {JCAP},
  volume = {02},
  pages = {007},
  year = {2025},
  eprint = {2310.15392},
  archivePrefix = {arXiv},
  primaryClass = {hep-ph},
  doi = {10.1088/1475-7516/2025/02/007}
}
\end{document}